%% file: main.tex
\documentclass[12pt]{article}
\input{notation.tex}

\usepackage{booktabs}
\usepackage{float}
\usepackage{placeins}
\usepackage{algorithm}
\usepackage{algorithmic}
\usepackage{multirow}
\newcommand{\C}{\mathcal{C}}

\usepackage{arydshln}
\begin{document}
\title{Graph Based, Adaptive, Multi Arm, Multiple Endpoint, Two Stage Design}
\author{ 
Cyrus Mehta$^{(+)}$, Ajoy Mukhopadhyay$ ^{(\dagger)} $, and Martin Posch$ ^{(\ddagger)} $\footnote{Correspondence: Martin Posch, Medical University of Vienna, martin.posch@meduniwien.ac.at}
\vspace*{.25in} \\ 
Cytel Corporation$ ^{(+)} $ \\  
Harvard T.H.Chan School of Public Health$^{(+)}$\\
 Novo Nordisk$^{(\dagger)}$\\
Medical University of Vienna$ ^{(\ddagger)} $\\ \\
}
\date{January 1, 2025}

\maketitle

\begin{center}
\Large
{\bf Abstract}
\normalsize
\end{center}
The graph based approach to multiple testing is an intuitive method that enables a study team to represent clearly, through a directed graph, its priorities for hierarchical testing of multiple hypotheses, and for propagating the available type-1 error from rejected or dropped  hypotheses to hypotheses yet to be tested. Although originally developed for single stage non-adaptive designs, we show how it may be extended to two-stage designs that permit early identification of efficacious treatments, adaptive sample size re-estimation, dropping of hypotheses, and changes in the hierarchical testing strategy at the end of stage one.  Two approaches are available for preserving the family wise error rate in the presence of these adaptive changes; the p-value combination  method, and the conditional error rate method. In this investigation we will present the statistical methodology underlying each approach and will compare the operating characteristics of the two methods in a large simulation experiment.


\section{Introduction} \label{sec:intro}
A common design issue in a confirmatory clinical trial comparing one or more treatment arms to a common control arm is how to perform hypothesis tests for more than one  efficacy endpoint while preserving strong control of the family wise error rate (FWER). For example in a double blind clinical trial of  patients with acute exacerbation of schizophrenia (ClinicalTrials.gov identifier NCT01469039) two doses of  the experimental drug ALK-9072  (350 mg/day or 500 mg/day) were compared to a matching control drug. The primary endpoint was the change from baseline in the PANSS total score at week 12. A key secondary endpoint was the change from baseline in the PANSS negative symptoms score. The trial would be considered a success if  efficacy could be demonstrated with either dose for the primary endpoint. Once efficacy is demonstrated for the primary endpoint, there interest in also filing a claim for the secondary endpoint. In settings such as this, a convenient way to conceptualize and display the study team's strategy for testing the various hypotheses in a hierarchical manner, while recognizing that some hypotheses are more important than others, is through a graph with weighted nodes representing the various hypotheses to be tested, and  directed edges for propagating the weights from rejected hypotheses to ones yet to be tested.

Graph based approaches were developed in ~\citep{bretz2009graphical,burman2009recycling} and \citep{maurer2013memory} to construct flexible sequentially rejective Bonferroni based tests in single stage designs. The graphs are a convenient way to define weights in a closed  testing procedure, \citep{marcus1976closed}, where each intersection hypothesis is tested by a weighted Bonferroni test. 
Subsequently \cite{bretz2011graphical} showed how, by separating the weighting strategy from the testing strategy,  more efficient weighted Simes and Dunnett type tests could replace Bonferroni tests when additional information about the correlation structure amongst the hypotheses was available. \cite{xi2017unified} extended the procedure to weighted mixed tests for settings where the correlation between test statistics is only known for subsets of test statistics. Next, these testing procedures were extended to  group sequential tests incorporating interim analyses. First, \cite{maurer2013multiple} considered the case where the correlation between test statistics is unknown. \cite{anderson2022unified} extended the group sequential tests to the weighted mixed tests for settings where for some (or all) test statistics the correlation is known. Finally, in the present paper, these approaches are extended to adaptive tests, allowing for adaptations such as the selection of hypotheses, or sample size reassessment after an unblinded interim analysis. Table \ref{tab:overview} gives an overview of the available fixed sample,  group sequential and adaptive group sequential  multiple testing procedures.

\begin{table}[htb]
 \caption{Overview of proposed fixed sample and group sequential multiple testing procedures and adaptive multiple testing procedures based on conditional error rates. * \cite{klinglmueller2014adaptive} mention early rejections as extension in the discussion.}
    \label{tab:overview}
    \centering \small
\begin{tabular}{|p{5.6cm}|>{\centering\arraybackslash}p{1.4cm}|c|c|>{\centering\arraybackslash}p{1.62cm}|>{\centering\arraybackslash}p{1.2cm}|c|}
\hline
 Method & early rejection & adaptive & weights & nonpara-metric & para-metric & mixed \\
 \hline
 \cite{bretz2009graphical,burman2009recycling,maurer2013memory} &&& \cellcolor{green!30}x & \cellcolor{green!30}x && \\
 \hline
 \cite{bretz2011graphical} &&& \cellcolor{green!30}x && \cellcolor{green!30}x & \\
 \hline
 \cite{xi2017unified} &&& \cellcolor{green!30}x & \cellcolor{green!30}x & \cellcolor{green!30}x & \cellcolor{green!30}x \\
 \hline
 \cite{anderson2022unified} & \cellcolor{green!30}x && \cellcolor{green!30}x & \cellcolor{green!30}x & \cellcolor{green!30}x & \cellcolor{green!30}x \\
 \hline
 \cite{koenig2008adaptive} && \cellcolor{green!30}x && & \cellcolor{green!30}x & \\
 \hline
 \cite{magirr2014flexible,ghosh2020adaptive,placzek2019conditional} & \cellcolor{green!30}x & \cellcolor{green!30}x && & \cellcolor{green!30}x & \\
 \hline
 \cite{posch2011type,klinglmueller2014adaptive,sugitani2016simple} & \cellcolor{green!30}$\text{x}^*$ & \cellcolor{green!30}x & \cellcolor{green!30}x & \cellcolor{green!30}x && \\
 \hline
 This paper & \cellcolor{green!30}x & \cellcolor{green!30}x & \cellcolor{green!30}x & \cellcolor{green!30}x & \cellcolor{green!30}x & \cellcolor{green!30}x \\
 \hline
\end{tabular}
   
\end{table}

We consider two approaches to construct weighted adaptive multiple testing procedures; one based on the combination test principle and the other based on the conditional error rate principle. In the first approach, adaptive p-value combination tests \citep{bauer1994evaluation}, in conjunction with closed testing,  are used to construct adaptive multiple testing procedures that control the family wise error rate by combining adjusted stage-wise p-values \citep{hommel2001adaptive,bretz2009adaptive}. We consider tests where the adjusted p-values are based on weighted Bonferroni tests if the correlation between test statistics is unknown \cite{bretz2009graphical}, weighted parametric tests if the correlation is known  \citep{bretz2011graphical},  or weighted mixed tests if  the correlation between only some of the test statistics is known \citep{xi2017unified}.

The second approach  is based on the conditional rejection principle \citep{proschan1995designed, posch1999adaptive, muller2004general}. For settings where the correlation between test statistics is known, this approach has been used to construct multiple testing procedures such as adaptive Dunnett tests \citep{koenig2008adaptive},  adaptive multi-arm multi-stage group sequential designs incorporating early stopping \citep{magirr2014flexible, ghosh2020adaptive}, and adaptive multiple testing procedures for clinical trials with subgroup selection \citep{placzek2019conditional}. However, these approaches use unweighted tests. For settings where the correlation is unknown, weighted adaptive tests based on partial conditional error rates have been proposed by \citep{posch2011type, klinglmueller2014adaptive,sugitani2016simple}, who generalize the adaptive, fixed sample graph based tests to adaptive tests for this scenario. 

In this paper, we extend both the combination approach and the conditional error approach to two stage group sequential weighted adaptive tests for scenarios where the correlation may be known between some test statistics and unknown in others. Non-negative weights can be defined by graphs \citep{bretz2009graphical, burman2009recycling, maurer2013memory} or, alternatively, arbitrary non-negative weights summing up to one (or a lower value) can be specified. Section~\ref{sec:graphspecific} is a quick review of the~\cite{bretz2009graphical} and \cite{bretz2011graphical} graph-based methodology,  applied to a single stage non adaptive design involving two treatment arms, a common control arm, and two endpoints. Section~\ref{sec:graphgeneral} provides the mathematical notation and technical details for a  general two stage graph based adaptive design involving $ k $ hypotheses. This section is split into two parts; Section~\ref{sec:Combo} covers the p-value combination method, while Section~\ref{sec:cer} covers the conditional error rate method. Within each of these sections is an illustrative example explaining how to perform the actual computations for the interim and final analyses.  Section~\ref{sec:simulations} presents the results of a simulation study  that demonstrates FWER control and compares the disjunctive and conjunctive power of the two methods over a range of scenarios and decision rules for selecting hypotheses at the end of stage one. In Section~\ref{sec:conclusions} we review our major findings, one of which is that the conditional error approach outperforms  the combination approach with respect to both disjunctive and conjunctive power across all scenarios and decision rules considered. We discuss the role that consonance might be playing in this important result  We end with some suggestions for further work.

The flexibilty to adaptively alter the future course of a confirmatory clinical trial while preserving its FWER has evolved considerably since the early days of the two arm group sequential design where early stopping for efficacy or futility were the only options. The present work incorporates all the important subsquent advances including sample size re-estimation, multiple treatment arms, multiple endpoints, treatment or endpoint selection, and graph based weighting strategies for hypothesis testing. The  incorporation of  the graph based methodology  into the multi-arm group sequential framework requires careful explanation, which has added to the length of the paper and the complexity of the notation.

\section{Non-Adaptive  Graph Based Closed Testing } \label{sec:graphspecific}
Consider the schizophrenia trial introduced in Section~\ref{sec:intro}. Let $ \mu_{0, \mbox{tot}}$ be the mean week-12 change from baseline in the total PANSS score for the control  arm and $ \mu_{0, \text{neg}} $ be the mean week-12 change from baseline in the negative symptoms PANSS score for the control arm. Similarly let $ \mu_{500, \text{tot}} $ be the mean week-12 change from baseline in the total PANSS score for the 500 mg  arm and $ \mu_{500, \text{neg}} $ be the mean week-12 change from baseline in the negative symptoms PANSS score for the 500 mg  arm. Finally let $ \mu_{350, \text{tot}} $ be the mean week-12 change from baseline in the total PANSS score for the 350 mg  arm and $ \mu_{350, \text{neg}} $ be the mean week-12 change from baseline in the negative symptoms PANSS score for the 350 mg  arm. Then $ \delta_{1}= \mu_{500, \text{tot}} - \mu_{0, \text{tot}} $ is the treatment effect of the high dose arm for the primary endpoint, $ \delta_{2}= \mu_{350, \text{tot}} - \mu_{0, \text{tot}} $ is the treatment effect of the low dose arm for the primary endpoint, $ \delta_{3}= \mu_{500, \text{neg}} - \mu_{0, \text{neg}} $ is the treatment effect of the high dose arm for the secondary endpoint and $ \delta_{4}= \mu_{350, \text{tot}} - \mu_{0, \text{tot}} $ is the treatment effect of the low dose arm for the secondary endpoint. 

The investigators are interested in testing each elementary hypothesis $ H_{j}\mbox{: } \delta_{j}=0 $ against the corresponding one-sided alternative hypothesis $ \delta_{j} >0 $, $ \text{for all }j \in I =\{1, 2, 3, 4\} $  while preserving strong control of the FWER at level $ \alpha $. This can be achieved by carrying out a closed test, which involves testing the  intersection hypotheses $ H_{J} = \cap_{j \in J}H_{j}$, for all subsets $ J \subseteq I $, with local level-$ \alpha $ tests. From the perspective of the investigators, however,  some hypotheses are more important than others. Therefore it is desirable to conduct weighted hypothesis tests with weights $ w_{j, J}, j \in J $ assigned to the components of each intersection hypothesis $ H_{J} $ for all $ J \subseteq I $. It is easy to show that there are $ |I|\times 2^{(|I|-1)}= 32$ individual weights to be assigned to the full closure tree of the four elementary hypotheses in $ I $.  Assigning these weights so that they reflect the study teams's relative priorities,  and  so that these priorities can be easily communicated to trial investigators, poses a challenge. The graph based weighting strategy proposed by~\cite{bretz2011graphical} is an extremely convenient way to achieve both goals.  Figure~\ref{fig:schizo1} is an example of such a strategy.
\begin{figure}[htp]
\begin{center}
\begin{tikzpicture}[scale=1.0]
\GraphInit[vstyle=Normal]
\SetVertexMath
\SetUpEdge[labelstyle = {draw}]
\Vertex[x=0, y=0]{H_3}
\Vertex[x=0, y=4]{H_1}
\Vertex[x=6, y=0]{H_4}
\Vertex[x=6, y=4]{H_2}
\node at (0, 4.85) {$ \frac{1}{2} $};
\node at (6, 4.85) {$ \frac{1}{2} $};
\node at (0, -0.75) {$ 0 $};
\node at (6, -0.75) {$ 0 $};
\tikzset{EdgeStyle/.style = {->,bend left}}
\Edge[label=1](H_4)(H_1)
\tikzset{EdgeStyle/.style = {->,bend right}}
\Edge[label=1](H_3)(H_2)
\tikzset{EdgeStyle/.style = {->}}
\Edge[label=$ \frac{1}{2} $](H_1)(H_3)
\Edge[label=$ \frac{1}{2} $](H_2)(H_4)
\tikzset{EdgeStyle/.style = {->,bend left}}
\Edge[label=$ \frac{1}{2} $](H_1)(H_2)
\tikzset{EdgeStyle/.style = {->,bend left}}
\Edge[label=$ \frac{1}{2} $](H_2)(H_1)
\end{tikzpicture}
\caption{}\label{fig:schizo1}
\end{center}
\end{figure}
 It consists of four nodes, one for each elementary hypothesis  $H_{j}, j \in I$, connected to each other in a specific way by directed edges. A nodal weight $ w_{j, I} $ is assigned to each node $ H_{j}, j \in I $. The edges of  the graph are specified through a transition matrix $ G=(g_{ij}) \leq 1 $ connecting the node $ H_{i} $ to the node $ H_{j}$ with edge weights $ 0 \leq g_{ij}, g_{ii} =0 $ and $ \sum_{j=1}^{|I|} g_{ij} \leq 1 $, for all $ i, j \in I $. In Figure~\ref{fig:schizo1} the nodal weights are,  $ w_{1, I} = w_{2, I} = 1/2 $ and $ w_{3, I} = w_{4, I} = 0 $. While these are the assigned weights for testing the  intersection hypothesis $ H_{I} = \cap_{j \in I } H_{j}$,  the choice of test  may differ, depending on what is known about the distribution of the marginal p-values $ p_{j}, j \in I$. For instance if the correlations amongst the p-values are completely unknown, one might choose the weighted Bonferroni  test, which rejects  $ H_{I} $ if $ \cup_{j \in I} [ p_{j} \leq w_{j, I} \alpha ]$. On the other hand if the correlations amongst the p-values are known, the  level-$ \alpha $ weighted Dunnett test,  which rejects $ H_{I} $ if  $ \cup_{j \in I} [p_{j} \leq c_{_{I}}w_{j, I} \alpha ]$, where $ c_{I} $ satisfies the relationship $ P_{H_{I}}\{ \cup_{j \in I} [P_{j} \leq c_{I} w_{j, I} \alpha] \}= \alpha $, would be more powerful. A testing strategy can also be constructed for the mixed case in which the correlations are known for some $ p_{j}, j \in I $, but not for others. In effect we have separated the process of assigning weights to the individual components of the intersection hypothesis $ H_{I} $ from the process of determining the appropriate weighted hypothesis test with which to test $ H_{I} $.  

Returning to Figure~\ref{fig:schizo1}, the transition matrix  is given by
\[
G =\left( \begin{array}{cccc}
0 & 1/2& 1/2& 0\\
1/2& 0& 0& 1/2\\
0& 1& 0& 0\\
1&0&0&0 \end{array}\right)
\]
where, as we shall see shortly, the edge weight $ g_{ij} $ represents the fraction of the nodal weight at node $ H_{i} $ that would be transferred to node $ H_{j} $ if $ H_{i} $ were removed from the graph. If $ g_{ij} =0 $ then  the nodes $ H_{i} $ and $  H_{j}  $ are not connected.  

\begin{figure}[htb]
\begin{center}
\begin{tikzpicture}[scale=0.85]
\GraphInit[vstyle=Normal]
\SetVertexMath
\SetUpEdge[labelstyle = {draw}]

\node at (3, 7.5)[font=\fontsize{12}{12}\selectfont]{(a) initial graph};
\Vertex[x=1.5, y=6] {H_1}
\Vertex[x=1.5, y=2] {H_3}
\Vertex[x=4.5, y=6] {H_2}
\Vertex[x=4.5, y=2]  {H_4}
\tikzset{EdgeStyle/.style = {->}}
\Edge[label=$ \frac{1}{2} $](H_1)(H_3)
\Edge[label=$ \frac{1}{2} $](H_2)(H_4)
\tikzset{EdgeStyle/.style = {->, bend left}}
\Edge[label=1](H_4)(H_1)
\tikzset{EdgeStyle/.style = {->, bend right}}
\Edge[label=1](H_3)(H_2)
\node at (1.5, 7)[font=\fontsize{11}{11}\selectfont]{$ \frac{1}{2} $};
\node at (1.5, 1)[font=\fontsize{11}{11}\selectfont]{0};
\node at (4.5, 7)[font=\fontsize{11}{11}\selectfont]{$ \frac{1}{2} $};
\node at (4.5, 1)[font=\fontsize{11}{11}\selectfont]{0};
\tikzset{EdgeStyle/.style = {->,bend left}}
\Edge[label=$ \frac{1}{2} $](H_1)(H_2)
\tikzset{EdgeStyle/.style = {->,bend left}}
\Edge[label=$ \frac{1}{2} $](H_2)(H_1)

\node at (8.5, 7.5)[font=\fontsize{12}{12}\selectfont]{(b) $ H_1 $ removed};
\Vertex[x=10, y=6] {H_2}
\Vertex[x=7, y=2] {H_3}
\Vertex[x=10, y=2] {H_4}
\tikzset{EdgeStyle/.style = {->}}
\Edge[label=1](H_3)(H_2)
\Edge[label=$ \frac{2}{3} $](H_2)(H_4)
\Edge[label=$ \frac{1}{2} $](H_4)(H_3)
\node at (10, 7)[font=\fontsize{11}{11}\selectfont]{$ \frac{3}{4} $};
\node at (7, 1)[font=\fontsize{11}{11}\selectfont]{$ \frac{1}{4} $};
\node at (10, 1)[font=\fontsize{11}{11}\selectfont]{0};
\tikzset{EdgeStyle/.style = {->,bend right}}
\Edge[label=$ \frac{1}{3} $](H_2)(H_3)
\tikzset{EdgeStyle/.style = {->,bend right}}
\Edge[label=$ \frac{1}{2} $](H_4)(H_2)

\node at (14, 7.5)[font=\fontsize{12}{12}\selectfont]{(c) $ H_1, H_2 $ removed};
\Vertex[x=12.5, y=2] {H_3}
\Vertex[x=15.5, y=2] {H_4}
\tikzset{EdgeStyle/.style = {->, bend left}}
\Edge[label=1](H_3)(H_4)
\Edge[label=1](H_4)(H_3)
\node at (12.5, 1)[font=\fontsize{11}{11}\selectfont]{$ \frac{1}{2} $};
\node at (15.5, 1)[font=\fontsize{11}{11}\selectfont]{$ \frac{1}{2} $};

\end{tikzpicture}
\caption{Impact of selective removal of $ H_{1} $ and $ H_{2} $ from original graph} \label{fig:schizo2}
\end{center}
\end{figure}
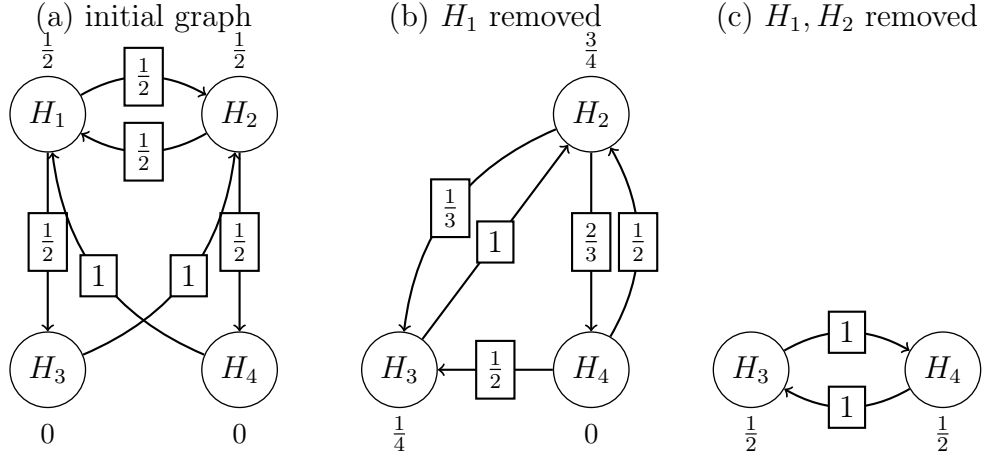

Although only the  weights for testing the intersection hypothesis $ H_{I} $ are displayed explicitly in Figure~\ref{fig:schizo1}, the weights for all the other intersection hypotheses $ H_{J}, J \subseteq I $ are actually embedded within it and can be extracted with the help of the transition matrix $ G $. For any fixed $ J \subseteq I  $, this is achieved by selectively removing each node from $ I \backslash J $,  re-connecting  the resulting loose edges, and updating the transition matrix, by means  of Algorithm 1 of~\cite{bretz2011graphical}. When all the nodes in $ I \backslash J $ have been thus removed, we are  left with a graph whose nodal values $ \{w_{j, J}, J \in J\} $ are the desired weights for the intersection hypothesis $ H_{J} $. For completeness, we reproduce Algorithm 1 of~\cite{bretz2011graphical} below.
\begin{algorithm}[htb]
 \caption{of~\cite{bretz2011graphical} for extracting the weights $ \{w_{j, J}, j \in J\} $ for  any intersection hypothesis $ H_{J}, J \subseteq I $, embedded in a  graph specified by initial weights $ \{w_{j, I}, j \in I\} $ and transition matrix $ G = (g_{ij}), i, j \in I$.   }\label{alg_graphweights}
\begin{algorithmic}
\STATE $I\leftarrow\{1,\ldots,k\}$
\WHILE{$I \backslash J \neq \emptyset$} 
\STATE $j\leftarrow \min{I\backslash J}$
\STATE  Update the graph:
\STATE $$I\to I \backslash \{j\}$$
\STATE\begin{equation*}
\label{rule1}
    w_{\ell, I} \leftarrow \left\{
        \begin{array}{ll}
            w_{\ell, I} + w_{j, I} g_{j\ell}, & \ell\in I \\
            0, & \mbox{ otherwise }
        \end{array}
    \right.
\end{equation*}
\STATE \begin{equation*}
\label{rule2}
    g_{\ell k} \leftarrow \left\{
        \begin{array}{ll}
            \frac{g_{\ell k}+g_{\ell j}g_{j k}}{1-g_{\ell j}g_{j \ell}}, & 
\ell,k\in I,\ell\not=k, g_{lj}g_{jl}<1 \\
            0, & \mbox{otherwise}
        \end{array}
    \right.
\end{equation*}
\ENDWHILE
\end{algorithmic}
\end{algorithm}
This algorithm must be repeated for all subsets $ J \in I $. By way of illustration,  Figure~\ref{fig:schizo2}(b) displays the resultant graph when the node $ H_{1} $ is removed from the initial graph and Figure~\ref{fig:schizo2}(c) displays the resultant graph when, in addition, the node $ H_{2} $ is removed.
When $ H_{1} $ is removed, the resultant graph provides the weights for the intersection hypothesis $ H_{J}$,  $ J=\{2, 3, 4\} $, as $ w_{2, J} = 3/4, w_{3, J}=1/4, w_{4, J}=0 $. When, additionally, $ H_{2} $ is removed, the resultant  graph provides the weights for the intersection hypothesis $ H_{J}, J =\{3, 4\} $, as $ w_{3, J}=1/2, w_{4, J}=1/2 $. Table~\ref{table:closureweights} displays all 32 weights for the weighting strategy implied by Figure~\ref{fig:schizo1}.

\begin{table}[htb]
\begin{center}
\caption{  Weighting Strategy implied by Figure~\ref{fig:schizo1} }\label{table:closureweights}
\begin{tabular}{|c|c|c|}\hline
Index &Intersection  & Weights \\

Set (J) &Hypothesis ($H_{J} $) & $ \{w_{j, J}, j \in J \}$ \\ \hline
$ \{1, 2, 3, 4\} $&$ H_{1}\cap H_{2} \cap H_{3} \cap H_{4 } $ & \{0.5, 0.5, 0, 0\} \\

$ \{2, 3, 4\} $&$ H_{2}\cap H_{3} \cap H_{4} $ &\{0.75, 0.25, 0\}\\
$ \{1, 3, 4\} $&$ H_{1}\cap H_{3} \cap H_{4} $ &\{0.75, 0, 0.25\}\\
$ \{1, 2, 4\} $&$ H_{1}\cap H_{2} \cap H_{4} $ &\{0.5, 0.5, 0\}\\
$ \{1, 2, 3\} $&$ H_{1}\cap H_{2} \cap H_{3} $ &\{0.5, 0.5, 0\}\\

$ \{3, 4\} $&$ H_{3}\cap H_{4} $ & \{0.5, 0.5\} \\
$ \{2, 4\} $&$ H_{2}\cap H_{4} $ & \{1, 0\} \\
$ \{2, 3\} $&$ H_{2}\cap H_{3} $ & \{0.75, 0.25\} \\
$ \{1, 4\} $&$ H_{1}\cap H_{4} $ & \{0.75, 0.25\} \\
$ \{1, 3\} $&$ H_{1}\cap H_{3} $ & \{1, 0\}\\
$ \{1, 2\} $&$ H_{1}\cap H_{2} $ & \{0.5, 0.5\} \\

$ \{4\} $&$ H_{4} $ & \{1\}\\ 
$ \{3\} $&$ H_{3} $ & \{1\}\\
$ \{2\} $&$ H_{2} $ & \{1\}\\
$ \{1\} $&$ H_{1} $ & \{1\}\\ \hline

\end{tabular}

\end{center}
\end{table}

We have shown that the graph $ \{w_{j, I}, j \in I \}$, together with the transition matrix $ G(I) = \{(g_{ij}), i, j \in I \}$, provides the weights for all the weighted intersection hypotheses constituting the closed test of the elementary  hypotheses $ H_{j}, j \in I $. Since different graphs would give rise to different weighting strategies, it may be asked why the graph displayed in Figure~\ref{fig:schizo1}, and not any other, should be used for testing the endpoints in the Schizophrenia trial. The reason is that Figure~\ref{fig:schizo1} reflects the priorities of the study team.  Figures~\ref{fig:schizo1} and \ref{fig:schizo2} bear a visual resemblance to the sequentially rejective hierarchical  testing scheme of~\cite{bretz2009graphical}. In that interpretation the total $ \alpha $ is split equally  between the two doses with a strict hierarchy within dose for testing the secondary endpoint only if the primary endpoint is first rejected. The edge values show that if a primary endpoint is rejected for a given dose, half its share of  $ \alpha $ is passed down to the corresponding secondary endpoint for that dose, and half to the primary endpoint for the other dose. If both primary and secondary hypotheses are rejected for a given dose, the accumulated $ \alpha $ is passed on for testing the primary endpoint of the other dose. When a hypothesis is rejected, the corresponding node is removed from the graph and the the graph is updated in accordance with Algorithm 1 of~\cite{bretz2009graphical} (which is different from Algorithm 1 of ~\cite{bretz2011graphical}). Testing continues until no hypothesis can be rejected. \cite{bretz2009graphical} have shown that the end result of the sequentially rejective testing is the same regardless of the order in which the nodal hypotheses are tested. 

This interpretation is a completely equivalent short cut for the weighted closed testing  procedure described here provided the weighted tests for the intersection hypotheses are consonant, as defined by~\cite{hommel2007powerful}. If the tests for the intersection hypotheses are not consonant, the weights of the intersection hypotheses elicited by Algorithm 1 of~\cite{bretz2011graphical} nevertheless result in a valid closed test, but the interpretation as a sequentially rejective hypothesis testing procedure no longer holds. One can, however, by analogy with the sequentially rejective testing procedure, interpret the graph as reflecting the priorities of the study team with the nodal weights of the initial graph representing the relative importance of the elementary hypotheses and the edge weighs representing the fraction of the nodal weight being transmitted from a source node to a successor node. It is thus both natural and convenient to incorporate and communicate the weighting strategy for a clinical trial with multiple endpoints through a graph. \cite{bretz2009graphical} provide examples of  several  graphs, each reflecting the different  priorities of the study team for the multiple endpoints being tested. 

So far we have reviewed the graph based multiple testing for a single-look non-adaptive trial. We will now generalize the testing procedure to two-stage adaptive designs in which there is $ \alpha $ spending for early identification of effective endpoints, endpoint selection, sample size re-estimation and alteration of the initial graph based on an unblinded analysis of the stage one data. 

\section{Generalization to Two-Stage Adaptive Designs} \label{sec:graphgeneral}
Assume that a set of $ k>1 $ one sided elementary null hypotheses $  H_{i} $, $ i \in I_{1} $, $ I_{1} = \{1, \ldots, k\} $ is to be tested with strong control of  the family wise error rate (FWER) at level $ \alpha $ over the two stages of an adaptive clinical trial. The hypotheses $ H_{i} $ may correspond to different treatments, subgroups, endpoints, or a combination of these.  In order to test all these hypotheses with strong FWER control,  it is necessary to implement a closed testing procedure. This is achieved by testing   all the intersection hypotheses $ H_{J}, J \subseteq I_{1} $ with local level-$ \alpha $ tests. Moreover, since some hypotheses might be more important than others, weights $ \{w_{j, J}, j \in J\} $ will be applied to the individual components of each intersection hypothesis $ H_{J} $. A natural and transparent way to specify these weights is through a graph-based weighting strategy as described in Section~\ref{sec:graphspecific}. Accordingly we assume that an initial graph defined by weights $ \{w_{j, I_{1}}, j \in I_{1}\} $ and transition matrix $ \{G(I_{1}) = (g_{ij}), i, j \in I_{1}\}$ has been specified.  All $ k\times2^{k} $ weights associated with the full closure tree of the $ k $ elementary hypotheses indexed by $ I_{1} $ can be extracted from the graph with the help of Algorithm~1 of \cite{bretz2011graphical}. 

Having specified the weights for all the intersection hypothesis tests $ H_{J}, J \subseteq I_{1} $, the next step is to specify the actual hypothesis tests  to which these weights will be applied. In an adaptive design, these hypothesis tests are specified in the planning phase, before any actual data from the trial are available. Then, at the end of stage one, on the basis of an unblinded interim analysis, these tests  may undergo adaptive modifications. The weighting stategy too may be modified through a revised graph, such that new weights $ \{\t w_{j, J}, j \in J \} $ are applied for stage two testing. Additional stage two data are then gathered and a final analysis is performed using all available data over both stages. Even though a closed test will be performed for the final analysis, these adaptive modifications could undermine strong FWER control unless appropriate adjustments are made to the testing procedure. In this paper we will present two adjustment methods, the p-value combination method, and  the conditional error rate method,  and will then compare their operating characteristics. The p-value combination method preserves the FWER by combining, for each intersection hypothesis, the p-value of the stage one data with the adapted p-value of the incremental stage two data using a pre-specified combination function. The conditional error rate method is based on a principle of  \cite{muller2001adaptive} which states that if a test is adapted at stage one, its probability of rejection under the null hypothesis, conditional on the observed stage one data, must not exceed the corresponding conditional rejection probability (CER) of the original unadapted test. In settings where the correlation between p-values of the individual hypotheses constituting an intersection hypothesis is unknown, the CER cannot be computed. In these cases we use an extension that controls the sum of partial conditional errors (PCER)  \citep{posch2011type,klinglmueller2014adaptive}.  We next discuss these two procedures in detail. Throughout, we will use the notational convention that $ p_{J, 2} $ is a p-value for the test of  $ H_{J} $ based on the cumulative data from stages one and two, whereas $ p_{J, (2)} $ is a p-value for the test of $ H_{J} $ based only on the incremental data obtained from stage two. Also $ P_{J, 2} $ and $ P_{J, (2)} $ are random variables whereas $ p_{J, 2} $ and $ p_{J, (2)} $ are specific values attained by them.

\subsection{The P-Value Combination Method} \label{sec:Combo}
In this method a combination test is pre-specified for each intersection hypothesis $ J \subseteq I_{1} $. At the end of stage one some intersection hypotheses are rejected. The trial proceeds to stage two, possibly after undegoing an adaptive modification, and  the remaining intersection hypotheses are tested.  The FWER of the individual hypotheses in $ I_{1} $ is preserved by a closed test. This is shown below in detail.

\subsubsection{Pre-planned Tests for the Intersection Hypotheses}
In the planning phase of  a two-stage adaptive trial, for each $ H_{J}, J \subseteq I_{1}$, we specify a level $ \alpha $ combination test $ \C_{J} $ with the decision function
\beqn
\varphi_{\C_{J}}(p_{J, 1}, p_{J, (2)}) =
    \begin{cases}
        1 & \text{if } p_{J, 1}\leq \alpha_{J, 1} \text{ or }\\
         & \quad p_{J, 1} > \alpha_{J, 1} \text{ and } \C_{J}(p_{J, 1}, p_{J, (2)})  \leq \alpha_{J, 2}\\ 
        0 & \text{ otherwise }
    \end{cases} \label{eq:Combo1}
\eeqn
where  $ p_{J, 1} $ and $ p_{J, (2)} $ are weighted adjusted  p-values for testing $ H_{J} $, based on the stage one and  incremental stage two data respectively, $ \C(p_{J, 1}, p_{J,(2)}) $ is a p-value combination function that combines the independent adjusted p-values of the two stages into a combined adjusted p-value incorporating the data from both stages, and $(\alpha_{J, 1}, \alpha_{J, 2}) $ are selected  so as to satisfy the level $ \alpha $ condition
\beqn
P_{H_{J}} (P_{J, 1} \leq \alpha_{1}) + P_{H_{J}} \{P_{J, 1} > \alpha_{J, 1} \text{ and } \C_{J}(P_{J,1}, P_{J, (2)}) \leq \alpha_{J, 2}\} = \alpha \ .  \label{eq:thelevelcondition}
\eeqn
In this investigation we will use the inverse normal p-value combination function defined by
\beqn
\C_{J}(p_{J, 1}, p_{J, (2)}) = 1-\Phi\{\nu_{J, 1} \Phi^{-1}(1- p_{J, 1}) + \nu_{J, 2} \Phi^{-1}(1-p_{J, (2)})\} \label{eq:thepvaluecombo}
\eeqn
in which $ \nu_{J, 1} $ and $ \nu_{J, 2} $ are pre-specified constants with $ \nu_{J, 1}^{2}+\nu_{J, 2}^{2} = 1 $, although other combination functions, (for example~\cite{bauer1994evaluation}) could be used as well. While the constants $ \nu_{J, 1} $ and $ \nu_{J, 2} $ could depend on $ J $, they are usually selected in proportion to the pre-planned sample sizes of the two stages of the trial. They must, however, remain fixed, even if the stage two sample size is adaptively altered.

The formula for computing the weighted adjusted p-values $ p_{J, 1}$ and $ p_{J, (2)} $ will depend on the available information about the correlations amongst the p-values of the elementary hypotheses constituting $ H_{J} $. Consider, for example, the intersection hypotheses displayed in Table~\ref{table:closureweights} for the schizophrenia trial described in Section~\ref{sec:graphspecific}. For some index sets $ J $ no information is available. This is the case, for example, for the intersection hypothesis $ H_{J} $ when $ J=\{1, 3\}$ in Table~\ref{table:closureweights}, since the correlation between the p-values for the primary and secondary endpoints of the schizophrenia trial is unknown. For other index sets the correlations amongst the p-values are completely known. This is the case, for example, for $ J=\{1, 2\} $ in Table~\ref{table:closureweights}. Here the correlation between the p-values of the two treatment arms of the schizophrenia trial can be deduced from the correlation between their inverse normal transformations, which depends on the allocation ratio of each treatment arm to the placebo arm.  Thus if we set $ Z_{1, 1} = \Phi^{-1}(1-P_{1, 1}) $ and $ Z_{2, 1} = \Phi^{-1}(1-P_{2, 1}) $ then, for balanced randomization, $ \text{corr}(Z_{1, 1}, Z_{2, 1}) = 0.5 $. And by the same reasoning  $ \text{corr}(Z_{1, (2)}, Z_{2, (2)}) = 0.5 $. Finally there is the mixed case in which the  correlations are known for some elements of $ J $ and unknown for others. This is the case, for example, for $ J=\{1, 2, 3 \}$ in Table~\ref{table:closureweights}. As a result, the weighted adjusted p-values for the intersection hypotheses $ H_{J}, J \subseteq I_{1} $ are of three types; nonparametric, parametric and mixed.
\begin{description}
\item[Nonparametric] If the correlations amongst the elementary p-values constituting the intersection hypothesis $ H_{J} $ are all unknown, compute the  weighted Bonferroni adjusted p-value from the stage one data as
\[
p_{J, 1} =\min \left(1,  \min_{j \in J}\{\frac{p_{j, 1}}{w_{j, J}}\}\right) \ ,
\]
and from the incremental stage two data as
\[
p_{J, (2)} =\min \left(1,  \min_{j \in J}\{\frac{p_{j, (2)}}{\tilde w_{j, J}}\}\right) \ .
\]

\item[Parametric] If the correlations amongst the p-values constituting the intersection hypothesis $ H_{J} $ are all known, compute the weighted parametric adjusted p-value from the stage one data as
\[
p_{J, 1} = P_{H_{J}} \left(\min_{j \in J}  \{\frac{P_{j, 1}}{w_{j, J}}\} \leq \min_{j \in J} \{\frac{p_{j, 1}}{w_{j, J}}\}\right) \ , 
\]
and from the incremental stage two data as
\[
p_{J, (2)} = P_{H_{J}} \left(\min_{j \in J}  \{\frac{P_{j, (2)}}{\t w_{j, J}}\} \leq \min_{j \in J} \{\frac{p_{j, (2)}}{\t w_{j, J}}\}\right) \ , 
\]
In the above formulae if $ w_{j, J}$ or $ \t w_{j, J} $ is zero for any $ j \in J $,  the corresponding weighted p-value term is dropped from both sides of the inequality.
\item[Mixed] If the correlations amongst the elementary p-values constituting the intersection hypothesis $ H_{J} $ are  unknown, for some $ j \in J $ and known for others, partition $ J $ into $ l $ distinct index sets   $J=\cup_{h=1}^{l}J_{h} $ such that the correlations amongst the elementary p-values constituting each index set $ J_{h}, h=1, \ldots l, $ are all known. Then the weighted adjusted p-value for the mixed case is

\[
p_{J, 1} =\min
\left\{1, 
\min_{h=1, \dots, l}
\left[\frac{P_{H_{J_{h}}} \left(\min_{j \in J_{h}}  \{\frac{P_{j, 1}}{w_{j, J}}\} \leq \min_{j \in J_{h}} \{\frac{p_{j, 1}}{w_{j, J}}\}\right)}
{\sum_{j \in J_{h}}w_{j, J}}
\right]
\right\} 
 \label{eq:pvmixed1} \ ,
\]
based on the stage one data, and is
\[
p_{J, (2)} =\min
\left\{1, 
\min_{h=1, \dots, l}
\left[\frac{P_{H_{J_{h}}} \left(\min_{j \in J_{h}}  \{\frac{P_{j, (2)}}{\t w_{j, J}}\} \leq \min_{j \in J_{h}} \{\frac{p_{j, (2)}}{\t w_{j, J}}\}\right)}
{\sum_{j \in J_{h}}\t w_{j, J}}
\right]
\right\} 
 \label{eq:pvmixed2} \ ,
\]
based on the incremental stage two data. Note  that one or more of the distinct subsets $ J_{h} $ partitioning the index set $ J $ can be singletons. The above formulae nevertheless remain applicable with the understanding that if  the weight $ w_{j, J}$  or $ \t w_{j, J} $ is zero for any singleton, that term it is dropped from the summation. 
\end{description}

\subsubsection{Closed Testing at the Interim and Final Analyses} \label{sec:pvcomboclosed}
Closed testing requires that every intersection hypothesis $ H_{J}, J \subseteq I_{1} $ should be tested with a local level-$\alpha $ test. At the interim analysis, after the stage one data have been observed, all the intersection hypotheses $H_J, J \subseteq I_{1}$,  that satisfy $p_{J,1}\leq \alpha_{J,1}$ are rejected.   For all $ J \subseteq I_{1} $, let $ \P(J) $ denote the set of all non-empty subsets of  $J$. That is, $ \P(J) $ is the power set of $J $.  Then $\mathcal{J}^r = \{J \in \P(I_{1})|p_{J,1}\leq\alpha_{J,1} \}$ denotes the  index set of rejected  intersection hypotheses. Since the status of  these intersection hypotheses has been determined at stage one,  they  need not be considered for further testing at stage two. It remains to determine, at stage two, the status of the remaining intersection hypotheses, indexed by $\mathcal{J}^+=\mathcal{P}(I_1)\backslash \{\mathcal{J}^r\}$, for which no test decision was made at the end of stage one. However, stage two data might not be available for computing adjusted stage two p-values $ p_{J,( 2)} $ for all the $ H_{J}$ with  $J \in \j^{+} $. To  see this we first observe that all individual hypotheses $H_i, i\in I_1$, for which all intersection hypotheses $H_J, J \subseteq I_{1}$ with $i\in J$, are in $\mathcal{J}^r$, are rejected by the closed test. Denote the index set containing these individual hypotheses by $ I_{1r} \subseteq I_{1}$. These hypotheses, having demonstrated efficacy at stage one itself, need not be tested again at stage two.  Therefore stage two data would be available, potentially, only for testing  the remaining individual hypotheses $ I^{*}_{1} = I_{1} \backslash \{ I_{1r}\}$. In a adaptive trial, however, there is also the option to drop additional elementary hypotheses from $ I_{1}^{*} $. Suppose then that, after an unblinded examination of the stage one data, it is decided to select only a subset $ I_{2} \subseteq I_{1}^{*} $ for stage two testing and to drop the remaining $ I^{*}_{1} \backslash I_{2} $ hypotheses from further consideration without gathering their stage two data.  The rationale for dropping  some hypotheses from $ I_{1}^{*} $ without stage two testing might be concerns about safety, efficacy, cost, enrollment, or  other  issues unrelated to formal statistical testing.  The adaptive procedure allows full flexibility to decide which hypotheses in $ I^{*}_{1} $ to drop and which ones to select for stage two testing. There is further flexibility to re-assess and alter, if necessary, the stage two sample size and the graph based representation of the priorities for testing the hypotheses indexed by $ I_{2} $. After making all these adaptive changes the trial proceeds to stage two for computing the adjusted p-values $ p_{J, (2)}, J \in \j^{+} $, with incomplete stage two data for some $ H_{J} $. Notwithstanding the missing data, these adjusted p-values must be specified since closed testing requires that $ H_{J}, J \in \j^{+} $  be tested with local level-$ \alpha  $ tests for all $ H_{J}, J \in \j^{+} $. To that end we partition $ \j^{+} $  into three non-overlapping subsets $ \j_{A}, \j_{B} $ and $ \j_{C} $ depending on the manner in which the adjusted p-values $ p_{J, (2)} $ are specified in each subset.
\begin{description} \label{page:jehplus}
 \item[$\mathcal{J}_A = \mathcal{P}(I_{2})\cap \mathcal{J}^{+}$:] Since the elementary hypotheses $H_i$, $i \in I_2$ will  be undergoing stage two testing, p-values $ p_{i, (2)}, i \in I_{2} $ will be available for each of them, and hence adjusted p-values for all intersection hypotheses $ H_{J}, J \in \mathcal{P}(I_{2}) $ can be derived.  

\item[$\mathcal{J}_B = \mathcal{P}(I_{1}^{*} \backslash I_{2}) \cap \mathcal{J}^{+}$:]  Since the elementary hypotheses $ H_{i} $, $ i \in I_{1}^{*} \backslash I_{2} $ will not have any stage two data, neither will the intersection hypotheses  $H_J, J\in  \mathcal{J}_{B}$. Therefore we set $p_{J, (2)}=1$ for all $ J \in \mathcal{J}_{B} $. 

\item[$\mathcal{J}_C =  \mathcal{J}^{+}  \backslash \{\mathcal{J}_{A}\cup \mathcal{J}_{B} \}$:] Since the intersection hypotheses $H_J, J\in \mathcal{J}_C$ belong neither to $ \mathcal{J}_{A} $  nor to $ \mathcal{J}_{B} $, it must be the case that  for all $ J \in \mathcal{J}_{C} $, $ J \cap I_{2} \neq J $ and $ J \cap I_{2} \neq \emptyset$. Therefore we will set the stage two p-value as
\begin{equation}
 \label{eq:ext}
p_{J, (2)}=p_{J\cap I_{2},(2)}
\end{equation}
for all $ J \in \mathcal{J}_{C} $.  Since $ J\cap I_{2}  \in \mathcal{J}_{A}$, these stage two p-values can be obtained from $ \mathcal{J}_{A} $. 
\end{description}

Adjusted p-values $ p_{J, 1} $ and $ p_{J,( 2)} $ are thus available for all $ J \in \mathcal{J}^{+} $ at the end of stage 2, so that  the combination test~(\ref{eq:Combo1}) can be evaluated for all  $ J \in  \{\mathcal{J}^{+} \cup \mathcal{J}^{r} \}= \mathcal{P}(I_{1})$. Finally, applying the closed testing principle,  a hypothesis $H_i, i\in I_2$, can be rejected at familywise error rate $\alpha$ if  all intersection hypotheses $H_{J},  J \subseteq I_{1} $ with $i\in  J$, are rejected with their respective level $\alpha$ combination tests. Thus for all $i\in I_{2}$ the decision function of the closed test of $ H_{i} $ is given by 
\beqn
\varphi_i=\min_{\{J \in \mathcal{P}(I_{1}),\ i\in J\}}\varphi_{C_J}(p_{J, 1},p_{J, (2)}) \ . \label{eq:decfuncombo}
\eeqn

\subsubsection{Illustrative Example of the P-Value Combination Method} \label{sec:pvexample}
Returning to the schizophrenia trial specified at the beginning of Section~\ref{sec:graphspecific}, suppose that the total sample size is 210 subjects with a balanced randomization of 70 subjects per arm. It is easy to show that with this sample size a single stage one sided level-0.025  Dunnett test has 89\% disjunctive power to detect a mean improvement of 10 units in the total PANSS score (given $ \sigma = 20 $). Now let  $ I_{1}=\{1, 2, 3, 4\} $ denote the index set of the four elementary hypotheses, and let the weighting strategy for testing all the intersection hypotheses $ H_{J}, J \subseteq I_{1} $ be represented graphically by Figure~\ref{fig:schizo1}, and in tabular form by Table~\ref{table:closureweights}. We will assume that there is an interim look after 50\% of the subjects have enrolled, and the trial spends the available one sided $ \alpha=0.025$ over the two stages of the trial in accordance with the  \cite{gordon1983discrete} O'Brien-Fleming (LDOF) type error spending function
\beqn
g(t, \alpha)=2 - 2 \Phi(\frac{z_{\alpha/2}}{\sqrt{t}}) \ , \label{eq:ldof}
\eeqn
where $ t  $ is the information fraction at the end of stage one. Since  $ t=0.5 $,  by~(\ref{eq:ldof}) the amount of $ \alpha $ available for testing each $ H_{J} $  at the end of stage one is $ \alpha_{J, 1}=0.00153 $. Suppose that at the end of stage one the marginal p-values for the elementary hypotheses are $ p_{1, 1} = 0.00045, p_{2, 1} = 0.0952, p_{3, 1} = 0.0225 $ and $ p_{4, 1} = 0.1104 $. These four marginal p-values,  are the building blocks for generating  the weighted adjusted p-value $ p_{J, 1} $, with the weights $ w_{J} = \{w_{j, J}, j \in J\} $ displayed in Table~\ref{table:closureweights},  for testing the intersection hypothesis $ H_{J}$ for all $ J \subseteq I_{1} $. Since $ \alpha_{J, 1}=0.00153 $, any $ H_{J} $ for which $ p_{J, 1} \leq 0.00153 $ will be rejected at stage one.

Table~\ref{table:closureweights2} extends Table~\ref{table:closureweights} by including two new columns containing, respectively, the weighted adjusted p-value and the corresponding adjustment method for all the intersection hypotheses $ H_{J}, J \subseteq I_{1} $ derived from $ p_{1,1}=0.00045, p_{2, 1}=0.0952, p_{3,1}=0.0225, p_{4, 1}=0.1104 $.

\begin{table}[htb]
\begin{center}
\caption{  Weighted adjusted p-values based on stage one marginal p-values. Note that for the computation of the adjusted p-values only the correlation between test statistics of hypotheses $H_{j} $ with $ w_{j, J}=0 $  needs to be known. }\label{table:closureweights2}
\begin{tabular}{|c|c|c|c|c|}\hline
Index &Intersection  & Weights&Adjusted p-value & Adjustment \\
Set (J) &Hypothesis ($H_{J} $) & $ \{w_{j, J}, j \in J \}$ & $ p_{J, 1} $  & Method\\ \hline
$ \{1, 2, 3, 4\} $&$ H_{1}\cap H_{2} \cap H_{3} \cap H_{4 } $ & \{0.5, 0.5, 0, 0\} &0.00088&Parametric\\

$ \{2, 3, 4\} $&$ H_{2}\cap H_{3} \cap H_{4} $ &\{0.75, 0.25, 0\}&0.0900&Nonparametric\\
$ \{1, 3, 4\} $&$ H_{1}\cap H_{3} \cap H_{4} $ &\{0.75, 0, 0.25\}&0.0006&Nonparametric\\
$ \{1, 2, 4\} $&$ H_{1}\cap H_{2} \cap H_{4} $ &\{0.5, 0.5, 0\}&0.00088&Parametric\\
$ \{1, 2, 3\} $&$ H_{1}\cap H_{2} \cap H_{3} $ &\{0.5, 0.5, 0\}&0.00088&Parametric\\

$ \{3, 4\} $&$ H_{3}\cap H_{4} $ & \{0.5, 0.5\} &0.0410&Parametric\\
$ \{2, 4\} $&$ H_{2}\cap H_{4} $ & \{1, 0\}&0.0952& NA\\
$ \{2, 3\} $&$ H_{2}\cap H_{3} $ & \{0.75, 0.25\}&0.0900& Nonparametric\\
$ \{1, 4\} $&$ H_{1}\cap H_{4} $ & \{0.75, 0.25\}&0.0006& Nonparametric\\
$ \{1, 3\} $&$ H_{1}\cap H_{3} $ & \{1, 0\}&0.00045& NA\\
$ \{1, 2\} $&$ H_{1}\cap H_{2} $ & \{0.5, 0.5\} &0.00088&Parametric\\

$ \{4\} $&$ H_{4} $ & \{1\} &0.1104&NA\\ 
$ \{3\} $&$ H_{3} $ & \{1\} &0.0225&NA\\
$ \{2\} $&$ H_{2} $ & \{1\} &0.0952&NA\\
$ \{1\} $&$ H_{1} $ & \{1\} &0.00045&NA\\ \hline
\end{tabular}

\end{center}
\end{table}

An examination of Table~\ref{table:closureweights2} reveals that the weighted adjusted p-values of all the  intersection hypotheses $ H_{J} $ containing $ H_{1} $ satisfy $ p_{J, 1} \leq \alpha_{J, 1} = 0.00153 $. Therefore the elementary hypothesis $ H_{1} $ is rejected by the closed test at stage one. We may thus eliminate the node $ H_{1} $ from the initial graph and construct a new graph with updated weights and edges in accordance with Algorithm~1  as shown in Figure~\ref{fig:schizo2}(b).

Before proceeding to stage two, the trial sponsors re-consider their priorities in  light  of $ H_{1} $ having been rejected at stage one itself. One option would be to discontinue any further enrollement to the high dose arm and re-assign the 35 subjects who would have been randomised to that arm during stage two to the remaining two arms. This would increase the power for detecting efficacy in the low-dose arm for both primary and secondary endpoints. The sponsors conclude, however,  that from both a business and patient benefit perspective it is more desirable to continue enrolling patients to all three arms in stage two as originally planned, because of the possibility of claiming efficacy for the secondary endpoint, PANSS for negative symptoms, in the high dose arm. We note that such a strategy carries the regulatory risk of being required to test $ H_{1} $ again within the closed testing framework. There is, however, no statistical requirement to do so, and it is reasonable to expect that as long as the stage two results for $ H_{1} $ are qualitatively consistent with what was observed in stage one, there will not be  any regulatory insistance on performing a closed test of $ H_{1} $ a second time. It is thus decided to select all three hypotheses $ H_{2}, H_{3} $ and $ H_{4} $ for stage two testing so that $ I_{2} = \{2, 3, 4\} $.   While the adaptive methodology of  closed combination testing would permit an increase in the sample sizes  of the three treatment arms, and would also permit changes in the hierarchical testing strategy, currently reflected in the graph~\ref{fig:schizo2}(b), the sponsors have decided to proceed to stage two without making any adaptive changes to the initial design. Accordingly, the study proceeds to stage two with the graphical testing strategy depicted in Figure~\ref{fig:schizo2}(b).

Suppose the  unadjusted p-values, based on the incremental stage two data, are $ \{p_{2, (2)} = 0.1121, p_{3, (2)} = 0.0112 , p_{4. (2)} = 0.1153\} $ respectively for $ \{H_{2}, H_{3}, H_{4}\} $. We are thus required to test  the intersection hypotheses $ H_{J} $ for all $ J \in \j^{+} =  \{ \{2\}, \{3\}, \{4\}, \{2, 3\}, \{2, 4\}, \{3,4\},  \{ 2, 3, 4\}\}$. Multiplicity adjusted  weighted p-values induced by the above three unadjusted p-values are computed in exactly the same way as was done for the stage one p-values. Table~\ref{table:adjpval2} displays the weighted adjusted stage two p-values and indicates whether they belong to the parametric, nonparametric or mixed cases.

\begin{table}[htb]
\begin{center}
\caption{
Adjusted p-values from incremental stage two data for all intersection hypotheses  $H_{J}, J \in \j^{+}  $ } \label{table:adjpval2}
\begin{tabular}{|c|c|c|c|c|}\hline
Index Set &Intersection Hypothesis & Weights & Adjusted p-value & Adjustment\\ 
$ J \in \j^{+}$  & $ H_{J} $ & $ \{w_{j, J}, j \in J\}$&$ p_{J, (2)} $ & Method \\ \hline
\{2, 3, 4\} & $ H_{2} \cap H_{3} \cap H_{4} $ & 0.75, 0.25, 0 & 0.0448 & Nonparametric \\ 
\{3, 4\} & $ H_{3} \cap H_{4} $ & 0.5, 0.5 & 0.0209 & Parametric\\ 
\{2, 4\} & $ H_{2} \cap H_{4} $ & 1.0, 0 & 0.1121 & NA\\
\{2, 3\} & $ H_{2} \cap H_{3} $ & 0.75, 0.25 & 0.0448 & Nonparametric \\
\{4\} & $ H_{4} $ & NA & 0.1153& NA \\
\{3\} & $ H_{3} $&NA & 0.0112 & NA\\
\{2\}&$ H_{2} $&NA&0.1121&NA \\  \hline

\end{tabular}

\end{center}
\end{table}

To complete the final analysis, the adjusted p-values of the two stages are combined with the pre-specified inverse normal combination function
\[
\C_{J}(p_{J, 1}, p_{J, (2)}) = 1- \Phi[\sqrt{\frac{1}{2}} \Phi^{-1}(1-p_{J, 1}) + \sqrt{\frac{1}{2}} \Phi^{-1}(1-p_{J, (2)}) ] 
\]
for all $ J \in \j^{+} $. The combined p-values are displayed in Table~\ref{table:pcombo3}.

\begin{table}[htb]
\begin{center}
\caption{
Final p-values, based on combining the adjusted p-values from the two stages, for all intersection hypotheses   $H_{J}, J \in \j^{+}  $  \label{table:pcombo3}
}
\begin{tabular}{|c|c|c|c|c|}\hline
Index Set & Intersection Hypothesis & \multicolumn{2}{c|}{Stage-wise Adjusted p-values }& Combined p-value \\ \cline{3-4}
$ J \in \j^{+} $ & $ H_{J} $ & $ p_{J, 1} $ & $ p_{J, (2)} $ & $ \C_{J}(p_{J, 1}, p_{J. (2)}) $ \\ \hline
\{2, 3, 4\} & $ H_{2} \cap H_{3} \cap H_{4} $ & 0.0900 & 0.0448 & 0.0158 \\
\{3, 4\} & $ H_{3} \cap H_{4} $ & 0.0410 & 0.0209 &  0.0038\\
\{2, 4\} & $ H_{2} \cap H_{4} $ & 0.0952 & 0.1121 & 0.0371 \\
\{2, 3\} & $ H_{2} \cap H_{3} $ & 0.0900 & 0.0448 & 0.0158 \\

\{4\} & $ H_{4} $ & 0.1104 &0.1153& 0.0433 \\
\{3\}& $ H_{3} $ & 0.0225& 0.0112 & 0.0012 \\
\{2\} & $ H_{2} $ & 0.0952 & 0.1121 & 0.0371\\ \hline

\end{tabular}

\end{center}
\end{table}

The hypothesis $ H_{J} $ is rejected at stage two if $ \C_{J}(p_{J, 1}, p_{J, 2}) \leq \alpha_{J, 2}$ where $ \alpha_{J, 2} $ satisfies~(\ref{eq:thelevelcondition}). It is easy to show by bivariate normal integration (see Remark 3 in the Supplementary Remarks of Section~\ref{sec:Combo} ) that $ \alpha_{J, 2} =  0.0245$. Then Table~\ref{table:pcombo3} shows that $ H_{3}, H_{2} \cap H_{3}, H_{3} \cap H_{4} $ and $ H_{2} \cap H_{3} \cap H_{4} $ are all rejected by their local level-$ \alpha $ tests. Thus $ H_{3} $ is rejected by a closed test at stage two. 

\subsection{The Conditional Error Rate Method} \label{sec:cer}
In this method a two-stage group sequential level-$ \alpha  $ test is pre-specified for each intersection hypothesis $H_{J},  J \subseteq I_{1} $. At the end of stage one a conditional error rate (CER), or a sum of partial conditional error rates (PCER),  is computed for each  intersection hypotheses not rejected early.  The trial proceeds to stage two, possibly after undergoing an adaptive modification. The unrejected intersection hypotheses are tested again, this time with modified critical cut-offs that preserve the CER or sum of PCERs, respectively. The FWER of the individual hypotheses in $ I_{1} $ is preserved by a closed test. This is shown below in detail.

\subsubsection{Pre-planned Tests Prior to Stage One}
Tests $ \varphi_{J} $ are pre-specified for all intersection hypotheses $H_{J}, J \subseteq I_{1} $. These tests differ depending on whether correlations amongst the individual p-values of the elementary hypotheses constituting $ H_{J} $ are all unknown (nonparametric case), all known (parametric case), or partially known (mixed case). Let $ p_{j, 1}$ be the p-value for the elementary hypothesis $ H_{j} $ based on stage one data and let 
\begin{equation}
    p_{j, 2}=1-\Phi(\sqrt{t}\, \Phi^{-1}(1-p_{j,1})+\sqrt{1-t}\,\Phi^{-1}(1-p_{j,(2)}) \label{eq:combpval2}
\end{equation}
denote the p-value for the elementary hypothesis $ H_{j} $ based on the data from both stages, for all $ j \in I_{1} $ in the pre-specified trial.  Here, $t$ denotes the preplanned information fraction of the first stage and $p_{j,(2)}$ the p-value of the hypothesis test computed based on the second stage data only.  Let $\{ w_{j, J}, j \in J \}$ be the graph based weights assigned to the components of  the intersection hypothesis $ H_{J} $ for each $ J \subseteq I_{1} $. 
\begin{description}
\item[Nonparametric] Suppose the joint distribution of the $ P_{j, i}$ and $ P_{j', i} $ is unknown for all $ j, j' \in J, j \neq j', i=1, 2 $, and that $ \alpha_{J, 1} $ has been pre-assigned as the amount of type-1 error to be spent for testing $H_{J}$ at the end of stage one, possibly through pre-specification of a standard group sequential error spending function. Then the first and second stage critical constants $ c_{J, 1} $ and $ c_{J,2} $ are obtained as solutions to the group sequential equations
\beqn
\sum_{j\in J} P_{H_j}\left\{
 P_{j, 1} \leq w_{j,J} c_{J, 1}
\right\}  =\alpha_{J, 1}
\label{eq:tgamma1}
\eeqn
\beqn
\sum_{j\in J} P_{H_j}\left\{
 [P_{j, 1} \leq w_{j,J} c_{J, 1}]  \mbox{ or } [ P_{j, 2} \leq w_{j,J}c_{J, 2}]\right\}=\alpha
\label{eq:tgamma2}
\eeqn
Since the individual p-values are assumed to be uniformly distributed and the weights $w_{j,J}$ sum up to one, $c_{J,1}=\alpha_{J,1}$.  Now we define the test statistic 
\beqn
\phi_{J} = \sum_{j\in J} I\left\{
 [p_{j, 1} \leq w_{j,J} c_{J, 1} ]  \mbox{ or } [ p_{j, 2} \leq w_{j,J} c_{J, 2}]\right\}\ ,
\label{eq:nparcer}
\eeqn
where $I\{\cdot\}$ denotes the indicator function. Note that $\phi_{J}$ corresponds to the number of rejections in the intersection hypothesis test of $H_J$.
Finally, the decision function $\varphi_J=\min(1,\phi_J)$ defines a conservative two-stage  level $ \alpha $ test  of $ H_{J} $, since 
$$ 
E_{H_{J}}(\varphi_{J}) \leq E(\phi_J) =\sum_{j\in J} E_{H_{j}} I \left\{[P_{j, 1} \leq w_{j,J} c_{J, 1}]  \mbox{ or } [ P_{j, 2} \leq w_{j,J}c_{J, 2}]\right\} = \alpha \ .
$$ 

\item[Parametric] Suppose the joint distribution of $P_{j, i}$ and $P_{j', i}$ is known for all $ j, j' \in J, j \neq j', i=1, 2 $. As in the nonparametric case, let $ \alpha_{J, 1} $ denote the amount of type-1 error to be spent for testing $H_{J}$ at the end of stage one. The critical constants $ c_{J, 1} $  and $ c_{J, 2} $ are evaluated as solutions to the multi-arm group sequential equations
\beqn
P_{H_{J}} \{\cup_{j \in J} [P_{j, 1} \leq w_{j, J} c_{J, 1}]\} = \alpha_{J, 1} \label{eq:dunpar1} \ ,
\eeqn
\beqn
P_{H_{J}}\{ \cup_{j \in J} [P_{j, 1} \leq w_{j, J}c_{J, 1} ]\text{ or } \cup_{j \in J} [P_{j, 2} \leq w_{j, J} c_{J, 2}]\} = \alpha \label{eq:dunpar2} \ .
\eeqn
These computations could be performed with the help of the R function {\tt pmvnorm} belonging to the library {\tt mvtnorm}.  The decision function
\beqn
\varphi_{J} = I \left\{ \cup_{j \in J} [p_{j, 1} \leq w_{j, J} c_{J, 1}]  \mbox{ or}  \cup_{j \in J} [p_{j, 2} \leq w_{j, J} c_{J, 2}]
\right\}\label{eq:parcer}
\eeqn
defines an exact two-stage level-$ \alpha $ test of  $ H_{J} $. That is, $ E_{H_{J}}(\varphi_{J}) = \alpha $. Note that in the above evaluation of the critical constants $ c_{J, 1} $ and $ c_{J, 2} $ all probability expressions in which $ w_{j, J}= 0 $ will be dropped. 

\item[Mixed] Suppose that for $ i=1, 2 $, the joint distribution of  $ P_{j, i}$ and $P_{j', i} $ is known for some $ j, j' \in J, j \neq j $, and unknown for others.  Let $ J $ be partitioned into $ l $ distinct index sets 
\beqn J = \cup_{h=1}^{l} J_{h} \label{eq:Jh}
\eeqn
such that the correlations amongst the elementary p-values constituting each subset $ J_{h}$ are known. Note some of these distinct subsets partitioning $ J $ can be singletons, implying that the correlations between their p-value and the  p-values belonging to any other subset are unknown. Furthermore, note that the $J_h$ depend in addition to $h$ also on the set $J$ under consideration. For notational convenience, however, the dependence of these sets on $J$ has been suppressed. Let $ \alpha_{J, 1}$ denote the amount of type-1 error allowable at stage one. Then the critical constants $c_{J,1},c_{J,2}$  are solutions to the group sequential equations
\[
\sum_{h=1}^l P_{H_{J_{h}}} \{\cup_{j \in J_{h}} [P_{j, 1} \leq w_{j, J} c_{J, 1}]\} = \alpha_{J, 1}\ ,
\]
\[
\sum_{h=1}^l P_{H_{J_{h}}}\{ \cup_{j \in J_{h}} [P_{j, 1} \leq w_{j, J}c_{J, 1} ]\text{ or } \cup_{j \in J_{h}} [P_{j, 2} \leq w_{j, J} c_{J, 2}]\} = \alpha \ . 
\]
To construct the test for $H_J$, we compute the test statistic
\beqn
\phi_{J} = \sum_{h=1}^{l} 
I\{
 \cup_{j \in J_{h}} [p_{j, 1} \leq w_{j, J} c_{J, 1}]  \mbox{ or}  \cup_{j \in J_{h}} [p_{j, 2} \leq w_{j, J} c_{J, 2}]\}
\label{eq:premixedtest}
\eeqn 
for the test of $H_J$.  Finally, $\varphi_J=\min(1,\phi_J)$ defines a conservative two-stage level $ \alpha $ test  of $ H_{J}$, because $E_{H_{J}}(\varphi_{J}) \leq E_{H_J}(\phi_J)=\alpha$. It is easy to show that the mixed case specializes to the nonparametric case if  $ J_{h}, h=1,  \ldots, l $ are all singletons, and specializes to the parametric case if $l=1 $.
\end{description}

\subsubsection{Adjustments to Pre-planned Tests due to Stage Two Design Adaptations}
At the end of stage one we identify the sets $ \j^{r}, I_{r1}, I_{1}^{*}$ and $ \j^{+} $, exactly as defined in Section~\ref{sec:pvcomboclosed}. Additionally the stage two portion of the trial may be adapted by selecting a subset $ I_{2} \subseteq I_{1}^{*} $ of the elementary  hypotheses that remain to be tested, altering the sample sizes or  allocation ratios of the treatment arms, and changing the weighting strategy for stage two through a revised graph. If the study design undergoes any of these  modifications the modified test $ \t \varphi_{J} $  may no longer satisfy the level-$ \alpha $ condition 
\beqn
E_{H_{J}} (\t \varphi_{J}) \leq \alpha \ . \label{eq:levalfmod}
\eeqn
It is possible, however, to  satisfy (\ref{eq:levalfmod}) by imposing on each $ \t \varphi_{J} $, a constraint  derived from a generalization of the conditional error rate principle due to~\cite{muller2001adaptive}, \cite{muller2004general}.  The constraint depends on the stage one data, represented by $ \chi_{1} $, and differs depending on whether the test for $ \t \varphi $ is nonparametric, parametric or mixed.

\begin{description}
\item[Nonparametric] Define
\beqn
B_{J}(\chi_{1}) = \sum_{j \in J} P_{H_{j}}\{ P_{j, 2} \leq w_{j,J} c_{J,2} | p_{j, 1}\}  \label{eq:bjchinparam}
\eeqn
as the sum of conditional probabilities for rejecting each individual hypothesis $ H_{j} $ as part of the intersection hypothesis $ H_{J} $, given the stage one data.We refer to each term in this sum as a partial conditional error rate (PCER). Then (\ref{eq:levalfmod}) is satisfied if  the adapted test $ \t \varphi_{J}$ is constrained by the condition
\beqn
E_{H_{J}} [\t \varphi_{J} | \chi_{1}] \leq B_{J}(\chi_{1}) \label{eq:pcernparam}
\eeqn
uniformly over all possible stage one outcomes $ \chi_{1} $. To see this note that expected value of the statistic (\ref{eq:nparcer}) can be written as
\[
E_{H_{J}}(\phi_{J}) = \sum_{j \in J} E_{H_{j}}\{I[P_{j, 1 } \leq w_{j, J}c_{J, 1} \text{ or } P_{j, 2} \leq w_{j, J} c_{J, 2} ]\}
\]
so that its conditional expection is
\[
E_{H_{J}}(\phi_{J} | \chi_{1}) = \sum_{j\in J} E_{H_{j}} I\{[P_{j, 2} \leq w_{j, J} c_{J, 2}] | p_{j, 1}\} = B_{J}(\chi_{1}) \ .
\]
It follows that
\[
E_{H_{J}}(\t \varphi_{J}) 
= 
E_{\chi_{1}} E_{H_{J}}[\t \varphi_{J} | \chi_{1}]
\leq 
E_{\chi_{1}} B_{J}(\chi_{1})
=
E_{\chi_{1}} E_{H_{J}}[ \phi_{J} | \chi_{1}] 
= 
E_{H_{J}} (\phi_{J}) = \alpha \ .
\]
We refer to (\ref{eq:pcernparam}) as the PCER condition. Note that, since $\varphi_J\leq\phi_J$, the pre-planned test satisfies the PCER condition. Thus if, after examining the stage one data, no adaptations are performed, the pre-planned testing procedure can be used without any error inflation. 

Operationally the PCER  condition implies that the stage two critical value $ \t c_{J, 2} $ for the adapted nonparametric test of $ H_{J} $ should satisfy the constraint
\beqn
\sum_{j \in J} P_{H_{j}} \{(\t P_{j, 2} \leq \t w_{j, J} \t c_{J, 2}) |( p_{j, 1}, j \in J)\} = B_{J}(\chi_{1}) \ , \label{eq:pcernparamo}
\eeqn
where (throughout) the `tilde'  above a symbol implies that the corresponding term may have been adaptively altered.

\item[Parametric] Define
\beqn
B_{J}(\chi_{1}) = P_{H_{J}} \{\cup_{j \in J} [P_{j, 2} \leq w_{j, J} c_{J, 2}] | (p_{j, 1}, j \in J)\} \label{eq:bjchiparam}
\eeqn
as the conditional error rate (CER), or conditional probability of rejecting $ H_{J} $ given the stage one data. Then (\ref{eq:levalfmod}) is satisfied if  the adapted test $ \t \varphi_{J}$ is constrained by the condition
\beqn
E_{H_{J}} [\t \varphi_{J} | \chi_{1}] \leq B_{J}(\chi_{1}) \label{eq:cerparam}
\eeqn
uniformly over all possible stage one outcomes $ \chi_{1} $. To see this note that because 
\[
E_{H_{J}}( \varphi_{J}| \chi_{1}) = E_{H_{J}} \{I [\cup_{j \in J} P_{J, 2} \leq w_{j, J} c_{J, 2}] | (p_{j, 1}, j \in J)\} = B_{J}(\chi_{1}) \ ,
\]
it follows that
\[
E_{H_{J}}(\t \varphi_{J}) 
= 
E_{\chi_{1}} E_{H_{J}}[\t \varphi_{J} | \chi_{1}]
\leq 
E_{\chi_{1}} B_{J}(\chi_{1})
=
E_{\chi_{1}} E_{H_{J}}[ \varphi_{J} | \chi_{1}] 
= 
E_{H_{J}} (\varphi_{J}) = \alpha \ .
\]
We refer to (\ref{eq:cerparam}) as the CER condition. It was proposed originally by \cite{muller2001adaptive}, \cite{muller2004general} for settings in which $ E_{H_{J}}(\varphi_{J} | \chi_{1}) $ could be evaluated. Notice that the CER condition is satisfied trivially by the pre-planned test since $ E_{H_{J}} (\varphi_{J}| \chi_{1}) = B_{J}(\chi_{1}) $. Thus, as with the nonparametric case, one is free to examine the stage one data and revert to the pre-planned test at stage two if there are no adaptations.  

Operationally the CER  condition implies that the stage two critical value $ \t c_{J, 2} $ for the adapted parametric test of $ H_{J} $ should satisfy the constraint
\beqn
P_{H_{J}} \{\cup_{j \in J} [\t P_{j, 2} \leq \t w_{j, J} \t c_{J, 2}] | (p_{j, 1}, j \in J)\} = B_{J}(\chi_{1}) \ . \label{eq:cerparamo}
\eeqn

\item[Mixed] Partition of the index set $J$ as shown in \eqref{eq:Jh} and compute, for each $  h=1, 2, \ldots l$, 
\[
P_{H_{J_{h}}} \{[\cup_{j \in J_{h}}(P_{j, 2} \leq w_{j, J}c_{J, 2})]| (p_{j, 1}, j \in J_{h}) \} , 
\]
 the conditional probability to reject any hypothesis $ H_{j}, j \in J_{h} $ in the test of $H_J$. Define
\beqn
B_{J}(\chi_{1}) = \sum_{h=1}^{l} P_{H_{J_{h}}} \{[\cup_{j \in J_{h}}(P_{j, 2} \leq w_{j, J}c_{J, 2})]| (p_{j, 1}, j \in J_{h}) \} \label{eq:bjchimixed}
\eeqn
as the sum of the above conditional rejection probabilities. We refer to each term in this sum as a partial conditional error rate (PCER). It is in fact the conditional error rate for the parametric test of $ H_{J_{h}} $, where $ H_{J_{h}} $ is a component of the mixed case intersection hypothesis $ H_{J} $. Then (\ref{eq:levalfmod}) is satisfied if  the adapted test $ \t \varphi_{J}$ is constrained by the condition
\beqn
E_{H_{J}} [\t \varphi_{J} | \chi_{1}] \leq B_{J}(\chi_{1}) \label{eq:pcermixed}
\eeqn
uniformly over all possible stage one outcomes $ \chi_{1} $.  To see this note that expected value of the statistic (\ref{eq:premixedtest}) can be written as
\[
E_{H_{J}}(\phi_{J}) = \sum_{h=1}^{l}E_{H_{J_{h}}} \{I\{\cup_{j \in J_{h}}[P_{j, 1} \leq w_{j, J} c_{J, 1}] \text { or } [P_{j, 2} \leq w_{j, j} c_{J, 2}]\}
\]
so that its conditional expectation is
\[
E_{H_{J}}(\phi_{J}|\chi_{1}) = \sum_{h=1}^{l} E_{H_{J_{h}}}
\left\{
I \left[
\cup_{j \in J_{h}} (P_{j, 2} \leq w_{j, J} c_{J, 2}) | (p_{j, 1}, j \in J_{h})
\right]
 \right\}
=
B_{J}(\chi_{1}) \ .
\] 
It follows that
\[
E_{H_{J}}(\t \varphi_{J}) 
= 
E_{\chi_{1}} E_{H_{J}}[\t \varphi_{J} | \chi_{1}]
\leq 
E_{\chi_{1}} B_{J}(\chi_{1})
=
E_{\chi_{1}} E_{H_{J}}[ \phi_{J} | \chi_{1}] 
= 
E_{H_{J}} (\phi_{J}) = \alpha \ .
\]
Since $B_J(\chi_{1})$ is the sum of partial conditional error rates of parametric tests, we refer to (\ref{eq:pcermixed}) as the PCER condition for the mixed case. Notice that since $ \varphi_{J} \leq \phi_{J} $, the PCER condition is satisfied by the pre-planned test  Therefore, as with the nonparametric and parametric cases, so also for the mixed case the pre-planned testing procedure can be performed if, after examining the stage one data, there are no adaptations. 

Operationally the PCER  condition implies that the stage two critical value $ \t c_{J, 2} $ for the adapted nonparametric test of $ H_{J} $ should satisfy the constraint
\beqn
\sum_{h=1}^{l}  P_{H_{J_{h}}} \{ \cup_{j \in J_{h}} 
\t P_{j, 2} \leq \t w_{j, J} \t c_{J, 2})|( p_{j, 1}, j \in J_{h})\} = B_{J}(\chi_{1}) \ . \label{eq:pcermixedo}
\eeqn
If  the $ J_{h}, h=1, \ldots l, $ are all singletons, condition \eqref{eq:pcermixedo} reduces to (\ref{eq:pcernparamo}), the operational PCER condition for the nonparametric case, whereas if $ l=1 $, \eqref{eq:pcermixedo} specializes to (\ref{eq:cerparamo}) the operational CER condition for parametric tests. 
\end{description}

\subsubsection{Final Analysis at the End of Stage Two} \label{sec:cerfinal}
For the final analysis at the end of stage two, adapted tests $ \t \varphi_{J} $ satisfying the above PCER or CER conditions must be performed for all intersection hypotheses $ H_{J}, J \in \j^{+} $. Observe first that for the nonparametric and mixed case tests it is possible to have $ B_{J}(\chi_{1}) \geq 1 $, thereby implying that the PCER condition holds no matter how the test of $ H_{J} $ is adapted and regardless of the stage two data. In this case, therefore, one can choose $ \t \varphi_{J} = 1 $ as the adapted test and  reject $ H_{J} $ at stage one itself. For all other cases it is convenient to partition the set $ \j^{+} $ into three non-overlapping subsets,  $\j^{+} =  \j_{A} \cup \j_{B}  \cup  \j_{C} $, such that, depending on the availability of stage two data, $ \t \varphi_{J} $ is defined appropriately within each subset.

\begin{description}
 \item[$\mathcal{J}_A = \mathcal{P}(I_{2})\cap \mathcal{J}^{+}$:] Since stage two data will be available for testing the individual hypotheses $ \{H_{j},  j \in I_{2}\} $,  adapted tests $ \t \varphi_{J} $ can be defined for all $ J \in \j_{A} $, as we shall show below.

\item[$ \j_{B} = \P(I_{1}^{*}\backslash I_{2}) \cap \j^{+}$:] Since stage two data will not be available for testing the  individual hypotheses $\{ H_{j}, j \in I_{1}^{*} \backslash I_{2}\} $  we will set $ \t \varphi_{J} = 0 $ for all $ J \in \j_{B} $.

\item[$ \j_{C}=\j^{+} \backslash \{\j_{A} \cup \j_{B}\} $:] Consider any  $ J \in \j_{C} $. Since  $ J \notin \j_{A} $ and $ J \notin \mathcal{J}_{B} $, stage two data will be available for testing some individual hypotheses $ H_{j}, j \in J $ but not for others. In particular stage two data will be available only for testing the individual hypotheses $ H_{j}, j \in J\cap I_{2} $. Therefore the adapted test $ \t \varphi_{J} $ will depend on the stage two data only through the stage two cumulative p-values for  hypotheses $H_j$ with $j\in J\cap I_2$. 
\end{description}

We now show how, based on the above partitioning, the adapted stage two tests $ \t \varphi_{J} $ may be constructed for all $ J \in \j^{+} $ so as to satisfy the PCER condition (\ref{eq:pcernparamo}) if nonparametric, the CER condition (\ref{eq:cerparamo}) if parametric, and the PCER condition (\ref{eq:pcermixedo}) if mixed, thereby ensuring that in all cases $ E_{H_{J}}(\t \varphi_{J}) \leq \alpha $. In all cases the adapted p-value for testing $ H_{j} $ at stage two is computed as
\begin{equation}
  \t  p_{j, 2}=1-\Phi(\sqrt{\t t_j}\, \Phi^{-1}(1-p_{j,1})+\sqrt{1-\t t_j}\,\Phi^{-1}(1-\t p_{j,(2)}) \ , \label{eq:cerpval2}
\end{equation}
where $\t t_j$, $ \t p_{j,(2)} $ and $ \t w_{j, J} $ indicate the actual information fraction after sample size reassessment,  the incremental stage two p-value, and adapted graph based weights that may differ due to adaptation of the pre-specified design at the end of stage one. If there is no adaptation, $ \t p_{j, 2} $ reverts to (\ref{eq:combpval2}) the p-value for the pre-planned test at stage two. 

\begin{description}
\item[Nonparametric] If $ J \in \j_{A} $ 
\begin{equation}
\t \varphi_{J}  = \left\{
\begin{array} {ll} 
1 & \text{if } B_J \geq 1\\
1 & \text{if } \cup_{j \in J} \{\t p_{j, 2} \leq \t w_{j, J} \t   c_{J, 2} \} \text{ where } \t c_{J, 2} \text{ satisfies the PCER condition}  \\ \label{eq:nparfinal}
    &\quad \sum_{j \in J} P_{H_{j}}\{(\t P_{j, 2} \leq \t w_{j, J} \t c_{J, 2}) | (p_{j, 1}, j \in J)\} = B_{J}(\chi_{1}) \\
0 & \text{otherwise}
\end{array} 
\right.
\end{equation}
If $ J \in \j_{B}$, $ \t \varphi_{J} = 0 $. If $ J \in \j_{C} $, replace $ J $ with $ J\cap I_{2} $ in the construction of $ \t \varphi_{J}$ above, except in the indices of $\tilde  w_{j,J}, \tilde c_{J,2}$ and $B_J$, where the index $J$ remains unchanged.

\item[Parametric] If $ J \in \j_{A} $,
\beqn
\t \varphi_{J}  = \left\{
\begin{array}{ll}
1 & \text{if }\cup_{j \in J}\{ \t p_{j, 2} \leq \t w_{j, J} \t c_{J, 2}\} \mbox{ where } \t c_{J, 2} \text{ satisfies the CER condition } \\ 
   &	\quad P_{H_{J}} \{[\cup_{j \in J}(\t P_{j, 2} \leq \t w_{j, J} \t c_{J, 2})] | (p_{j, 1}, j \in J)\} = B_{J}(\chi_{1}) \\
	0 &  \mbox{ otherwise}  \label{eq:parfinal}
\end{array}
\right.
\eeqn
If $ J \in \j_{B}$, $ \t \varphi_{J} = 0 $. If $ J \in \j_{C} $, replace $ J $ with $ J\cap I_{2} $ in the  construction of $ \t \varphi_{J}$ above, except in the indices of $\tilde  w_{j,J}, \tilde c_{J,2}$ and $B_J$, where the index $J$ remains unchanged.

\item[Mixed] We use the partition of the index set $J$ defined in \eqref{eq:Jh}. Then, the mixed parametric  test of $ H_{J} $ is given by
\begin{equation}
\tilde{\varphi}_J = 
\begin{cases} 
1 & \text{if } B_J \geq 1, \\ 
1 & \text{if } \cup_{h=1}^l \cup_{j \in J_h} \{ \tilde{p}_{j,2} \leq \tilde{w}_{j,J} \tilde{c}_{J,2} \}, \\ 
  & \text{where } \tilde{c}_{J,2} \text{ satisfies the PCER condition} \\ 
  & \sum_{h=1}^l P_{H_{J_h}} \big\{ 
    \big[\cup_{j \in J_h} (\tilde{P}_{j,2} \leq \tilde{w}_{j,J} \tilde{c}_{J,2}) \big] 
    \big| (p_{j,1}, j \in J_h) \big\} = B_J(\chi_1), \\
0 & \text{otherwise}.
\end{cases} 
\label{eq:mixedfinal}
\end{equation}

If $ J \in \j_{B}$, $ \t \varphi_{J} = 0 $. If $ J \in \j_{C} $, replace $ J_{h} $ with $ J_{h}\cap I_{2} $ in the construction of $ \t \varphi_{J}$ above, except in the indices of $\tilde  w_{j,J}, \tilde c_{J,2}$ and $B_J$, where the index $J$ remains unchanged.  The mixed case test~(\ref{eq:mixedfinal}) is the most general way to test any adapted intersection hypothesis $ H_{J} $. It specializes to the nonparametric test~(\ref{eq:nparfinal}) if the $ \{J_{h}, h=1, 2, \ldots l\} $, are all singletons, and specializes to the parametric test~(\ref{eq:parfinal}) if $ l=1 $.
\end{description}
{\bf Supplementary Remarks}
\begin{enumerate}
\item In all three cases above it is assumed that if $ J \in \j_{C} $ no weight is given to dropped hypothesis, such that $\sum_{j\in J\cap I_2}\t w_{j,J}=1$, and $\t w_{j, J}=0$ for all $j \in J\setminus I_2$. 
\item For many settings, the conditional probabilities \eqref{eq:pcernparamo}, \eqref{eq:cerparamo} and \eqref{eq:pcermixedo} can be computed from multivariate normal distributions for z-statistics obtained by inverse normal transformations of corresponding p-value statistics. 
\end{enumerate}

\subsubsection{Illustrative Example of the Conditional Error Rate Method} \label{sec:cerexample}
We will repeat the analysis of the schizophrenia trial at one sided $ \alpha=0.025 $, this time by the CER method, keeping the weighting strategy and all other design parameters the same as in Section~\ref{sec:pvexample}.  The first step is to pre-specify the tests for all the intersection hypotheses $ H_{J}, J \subseteq I_{1} $ that would be performed if there were no adaptation at the end of stage one. For nonparametric tests (\ref{eq:nparcer}) this involves specifying the critical cut-off  values $ w_{j, J}c_{J, i} $, for corresponding elementary p-values $p_{j, i},  j \in J , i=1, 2$.   
We illustrate below with a couple of examples. 

Consider pre-specification of the test of $ H_{J} $ where $ J=\{1, 2, 3, 4\}$. From Table~\ref{table:closureweights2}, $ w_{1, J}=0.5, w_{2, J}=0.5, w_{3, J}=0, w_{4, J} = 0 $. Therefore the elementary p-values $ p_{3, i} $ and $ p_{4, i} $, having zero weights associated with them, will play no role in the test of $ H_{J} $.  Acceptance or rejection of $ H_{J} $ will depend solely on $ p_{1, i} $ and $ p_{2, i} $, $ i=1, 2 $. Since  $ \text{corr}(Z_{1-P_{1, i}}, Z_{1-P_{2, i}}) = 0.5 $, we are in the parametric setting and need to evaluate $ c_{J, 1} $ from (\ref{eq:dunpar1}) and $ c_{J, 2} $ from (\ref{eq:dunpar2}). We will be using the~\cite{gordon1983discrete} error spending function (\ref{eq:ldof}) and taking the interim look at information fraction $ t=0.5 $. Thus
\beqn
\alpha_{J, 1} = 2- 2\Phi(\frac{z_{\alpha/2}}{\sqrt{t}}) \ ,   \label{eq:ldoffun}
\eeqn
Solving (\ref{eq:dunpar1}) for the critical constant $ c_{J, 1} $ we have, 
\[
P_{J} \{\cup_{j \in \{1, 2\}} [P_{j, 1} \leq w_{j, J} c_{J, 1}]\} = 0.001525 
\]
whereupon $ c_{J, 1} = 0.001564$. Similarly, solving (\ref{eq:dunpar2}) for the critical constant $  c_{J, 2}$ we have
\[
0.001525 + P_{H_{J}}\{ \cap_{j \in J} [P_{j, 1} >  0.001564 w_{j, J}]\text{ and } \cup_{j \in J} [P_{j, 2} \leq w_{j, J} c_{J, 2}]\} = \alpha
\] 
whereupon $ c_{J, 2} =  0.02633$.
Thus the pre-specified p-value boundaries for $ \varphi_{J} $ in (\ref{eq:parcer}) are $ w_{1, J}c_{J, 1}=w_{2, J}c_{J, 1} = 0.000782$, $w_{3, J}c_{J, 1}=w_{4, J}c_{J, 1}=0$ at stage one, and $ w_{1, J}c_{J, 2}=w_{2, J}c_{J, 2} = 0.0132, w_{3, J}c_{J, 2}=w_{4, J}c_{J, 2}=0$ at stage two.

Next consider the pre-specification of the test of $ H_{J} $ where $ J=\{2, 3, 4\} $. From Table~\ref{table:closureweights2}, $ w_{2, J}=0.75, w_{3, J}=0.25, w_{4, J} = 0 $. Since $ w_{4, J} = 0 $, the elementary p-values $ p_{4, i}, i=1, 2,$ will play no role in the test of $ H_{J} $ and we need only concern ourselves with critical cut-off values for $ p_{2, i} $ and $ p_{3, i} $. Since the correlation between $ P_{2, i} $ and $ P_{3, i} $ is unknown, we are in the nonparametric setting. Solving the group sequential equation (\ref{eq:tgamma1}) for $ c_{J, 1} $ yields $ c_{J, 1} = \alpha_{J, 1} = 0.001525$. Therefore  the pre-specified stage one p-value boundaries  for $ \varphi_{J} $ in (\ref{eq:nparcer}) are $w_{2,J} c_{J, 1} = 0.001144, w_{3,J} c_{J, 1} = 0.000381   , w_{4,J} c_{J, 1} = 0$. Solving the group sequential equation (\ref{eq:tgamma2}) for $ c_{J, 2} $ yields $c_{J,2}=0.024409$. Thus the corresponding pre-specified stage two  p-value boundaries in (\ref{eq:nparcer}) are $w_{2,J} c_{J, 2} =0.0183, w_{3,J} c_{J, 2} = 0.00610, w_{4,J} c_{J, 1} = 0$. 

The weights $ \{w_{j, J}, j \in J \}$ and the pre-specified stage one and stage two p-value boundaries  are displayed in  Table~\ref{table:precerbdry} for all intersection hypotheses $ H_{J}, J \subseteq I_{1} $ .

\begin{table}[htb] \fontsize{10}{12}\selectfont
\begin{center}
\caption{Pre-specified Stage One and Stage Two Boundaries for CER Method  }\label{table:precerbdry}
\begin{tabular}{|c|c|c|c|c|} \hline
Intersection & Weights & \multicolumn{2}{c|}{P-value Boundaries} & Type\\ \cline{3-4}
Hypotheses $ (H_{J}) $ & $\{w_{j, J}, j \in J\}  $ & Stage One & Stage Two & of Test$ ^{(+)} $ \\ \hline
$ H_{1}\cap H_{2}\cap H_{3}\cap H_{4} $ &\{0.5, 0.5, 0, 0\} & \{0.000782, 0.000782, 0, 0 \}&\{0.0132, 0.0132, 0, 0\}&Parametric \\
$ H_{2} \cap H_{3}\cap H_{4} $ &\{0.75, 0.25, 0\} &\{0.00114, 0.000381, 0\}& \{0.0183, 0.00610, 0\}&Nonparametric \\
$ H_{1} \cap H_{3} \cap H_{4} $ & \{0.75, 0, 0.25\} & \{0.00114, 0,  0.000381\}&\{0.0183, 0, 0.00610\} & Nonparametric \\
$ H_{1}\cap H_{2} \cap H_{4} $ & \{0.5, 0.5, 0\} &\{0.000782, 0.000782, 0\} &\{0.0132, 0.0132, 0\} & Parametric \\
$ H_{1}\cap H_{2} \cap H_{3} $& \{0.5, 0.5, 0\} &\{0.000782, 0.000782, 0\} &\{0.0132, 0.0132, 0\} & Parametric \\
$ H_{3} \cap H_{4} $ & \{0.5, 0.5\} & \{0.000782, 0.000782\} & \{0.0132, 0.0132\} & Parametric \\
$ H_{2}\cap H_{4} $& \{1, 0\} & \{0.001525, 0\} & \{0.0245, 0\} & NA \\
$ H_{2} \cap H_{3} $&\{0.75, 0.25\} &\{0.00114, 0.000381\}&\{0.0183, 0.00610\} & Nonparametric \\
$ H_{1} \cap H_{4} $ & \{0.75, 0.25\} & \{0.00114, 0.000381\}&\{0.0183, 0.00610\}& Nonparametric\\
$ H_{1}\cap H_{3} $ & \{1, 0\} & \{0.001525, 0\} & \{0.0245, 0\} & NA \\
$ H_{1}\cap H_{2} $ & \{0.5, 0.5\} &\{0.000782, 0.000782\} &\{0.0132, 0.0132\} &Parametric \\
$ H_{4} $ & NA & 0.001525 & 0.0245 & NA\\
$ H_{3} $ & NA & 0.001525 & 0.0245 & NA\\
$ H_{2} $ & NA & 0.001525 & 0.0245 & NA\\
$ H_{1} $ & NA & 0.001525 & 0.0245 & NA\\  \hline
\end{tabular}
\end{center}
\end{table}

Suppose the stage one p-values for the elementary hypotheses in $ I_{1} $ are $ p_{1.1}=0.00045$, $ p_{2, 1}=0.0952 $, $ p_{3, 1}=0.0225 $, and  $p_{4, 1}= 0.1104 $. Applying these observations to the stage one p-value boundaries in Table~\ref{table:precerbdry} it is seen that every intersection hypothesis containing $ H_{1} $ is rejected whereas all other intersection hypotheses are retained. Therefore $ H_{1} $ is rejected under closed testing. Formally, the index set
$
\j^{r} = \{(1, 2, 3, 4), (1, 3, 4), (1, 2, 4), (1, 2, 3), (1, 4), (1, 3), (1, 2), (1)\} \ ,
$
$ I_{r1} =1 $, $ I_{1}^{*} = \{2, 3, 4\} $, and 
$
\j^{+}=\{(2, 3, 4), (3, 4), (2, 4), (2, 3), (4), (3), (2)\}\ .
$

In order to complete a level-$ \alpha $ closed test for all the elementary hypotheses $H_{j},  j \in I_{1} $ we are required to test all the intersection hypotheses $ H_{J}, J \in \j^{+} $ with local level-$\alpha$ tests at the end of stage two. Before proceeding to stage two, however, there is the option to make adaptive changes to the on going trial. We have indicated in Section~\ref{sec:cer} several ways in which the trial may be adapted including dropping of  hypotheses, sample size re-estimation and altering the testing strategy. Should any of these changes be implemented it will be necessary to replace the pre-specified tests $ \varphi_{J} $ with modified tests $ \t \varphi_{J} $ for all $ J \in \j^{+} $ such that the modified tests satisfy the required CER or PCER conditions.    Accordingly Table~\ref{table:cerlisting} displays $ B_{J}(\chi_{1}) $ values as computed by (\ref{eq:bjchinparam}) for the nonparametric tests and by  (\ref{eq:bjchiparam}) for the parametric tests,  for all the intersection hypotheses $ H_{J}, J \in \j^{+} $.

\begin{table}[htb]
\begin{center}
\caption{Required Values of $ B_{J}(\chi_{1}) $ for all $ J \in \j^{+}$ conditional on stage one p-values }\label{table:cerlisting}
\begin{tabular}{|l|l|l|} \hline
Intersection &Weights & \\
 Hypotheses ($ J \in \j^{+} $) &  $ \{w_{j, J}, j \in J\} $ & $B_{J}(\chi_{1}) $\\ \hline
$ H_{2}\cap H_{3} \cap H_{4} $ &\{0.75, 0.25, 0\} & $ 0.1117$ \\
$ H_{3} \cap H_{4} $ & \{0.5, 0.5\} & $ 0.1420 $ \\ 
$ H_{2} \cap H_{4} $ & \{1, 0\} & $ 0.0702$ \\
$ H_{2} \cap H_{3} $ & \{0.75, 0.25\} & $ 0.1117  $ \\
$ H_{4} $ & 1 & $  0.0594 $\\
$ H_{3} $ & 1 & $  0.2179 $\\
$ H_{2} $ & 1 & $  0.0702 $\\ \hline
\end{tabular}
\end{center}
\end{table}

\begin{figure}[htp]
\begin{center}
\begin{tikzpicture}[scale=1.0]
\GraphInit[vstyle=Normal]
\SetVertexMath
\SetUpEdge[labelstyle = {draw}]
\Vertex[x=0, y=0]{H_2}
\Vertex[x=6, y=0]{H_4}
\node at (0, -0.75) {$ \frac{1}{2} $};
\node at (6, -0.75) {$\frac{1}{2} $};
\tikzset{EdgeStyle/.style = {->,bend left}}
\Edge[label=1](H_2)(H_4)
\tikzset{EdgeStyle/.style = {->,bend left}}
\Edge[label=1](H_4)(H_2)
\end{tikzpicture}
\caption{ Graph of Stage Two Testing Strategy for Schizophrenia Trial using CER Method}\label{fig:rev2}
\end{center}
\end{figure}
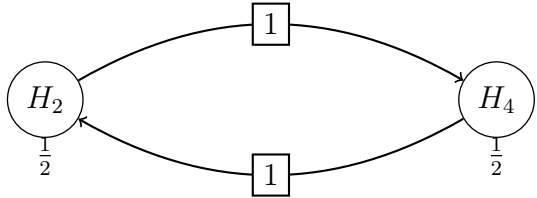

Suppose it is decided to drop $ H_{3} $, the secondary endpoint for the high dose arm,  on the grounds that, having  already rejected $ H_{1} $, the primary hypothesis for the high dose arm, it is preferable to now focus all remaining sample size resources only on the low dose arm. By this decision the additional 35 subjects who would have been randomized to the high dose arm at stage two  will now be  allocated to the low dose or control arms, leading to a second stage incremental sample size of $52$ and $53$ in the low dose and control group, respectively. The adapted information fraction is thus
\[
\t t = \frac{\left[     1/35 +1/35\right]^{-1}}{\left( \left[1/35+1/35\right]^{-1} + \left[1/52 +1/53    \right] ^{-1}     \right)         } = 0.4
\]
We are now left with only the two elementary hypotheses $ H_{2} $ and $ H_{4} $, indexed by $ I_{2} = \{2, 4\} $. Suppose that it is also decided to alter the testing strategy and  treat the two endpoints for the low dose arm as co-primaries, and to therefore assign equal weigh to $ H_{2} $ and $ H_{4} $. The graph representing the revised stage two testing strategy is displayed in Figure~\ref{fig:rev2}.

Because of the above adaptive changes at the end of stage one, the pre-specified p-value boundaries at stage two for the adapted tests $\t \varphi_{J},  J \in \j^{+} $ must be recomputed. For this purpose we first partition $ \j^{+} $ into the three non-overlapping subsets $ \j_{A} = \{(2), (4), (2, 4)\} $, $ \j_{B} = \{(3)\} $ and $ \j_{C} = \{(2,3,4), (3, 4), (2, 3)\} $ as defined  in Section~\ref{sec:cerfinal}. Then $ \t \varphi_{\{3\}} = 0$. The boundaries for the remaining hypotheses are displayed in Table~\ref{table:finalbdriesJA} for all $ J \in \j_{A} $ and in Table~\ref{table:finalbdriesJC} for all $ J \in \j_{C} $. The Table~\ref{table:finalbdriesJA} boundaries were evaluated in accordance with equation~(\ref{eq:pcernparamo}). The Table~\ref{table:finalbdriesJC} boundaries were evaluated in accordance with equation~(\ref{eq:pcernparamo}) after replacing each index set $ J $ with $ J\cap I_{2} $ in the summation, except in the indices of $ \t w_{j, J}, \t c_{J, 2} $ and $ B_{J} $

\begin{table}[htb]
\begin{center}
\caption{Re-computed stage two p-value boundaries for intersection hypotheses $ H_{J}, J \in \j_{A} $ }\label{table:finalbdriesJA}
\begin{tabular}{|c|c|c|c|} \hline
Index Set& Weights & Boundary & Boundary \\
$ J $ & $ \{\t w_{j, J}, j \in J\} $ & Type &   Values \\ \hline
 
\{2, 4\} & \{0.5, 0.5\} & Nonparam & $p_{2,2}\leq 0.0187    \text{ or } p_{4, 2} \leq  0.0187$\\ 

\{4\} & 1 & NA & $ p_{4, 2} \leq 0.0287 $ \\ 

\{2\} & 1 & NA & $ p_{2, 2} \leq 0.0274 $ \\ \hline

\end{tabular}
\end{center}
\end{table}

\begin{table}[htb]
\begin{center}
\caption{Re-computed stage two p-value boundaries for intersection hypotheses $ H_{J}, J \in \j_{C} $ }\label{table:finalbdriesJC}
\begin{tabular}{|c|c|c|c|c|} \hline
Index Set& Restricted Set&Weights & Boundary & Boundary \\
$ J $ &$ J \cap I_{2} $ &$ \{\t w_{j, J\cap I_{2}}, j \in J\cap I_{2}\} $ & Type &   Values \\ \hline

\{2, 3, 4\} &\{2, 4\}& \{0.5, 0.5\} & Nonparam & $p_{2,2}\leq 0.0253 \text{ or } p_{4, 2} \leq 0.0253$\\ 

$ \{3, 4\} $ & \{4\} & 1 & NA & $ p_{4, 2} \leq  0.0528 $ \\

$ \{2, 3\} $ & \{2\} & 1 & NA & $ p_{2, 2} \leq 0.0382 $ \\ \hline

\end{tabular}
\end{center}
\end{table}

The  trial  proceeds to stage two. Suppose that at the end of stage two the incremental p-values are
$\t p_{2,(2)}=0.0299, \t p_{4,(2)}=0.0586$
leading, in accordance with (\ref{eq:cerpval2}), to cumulative p-values over the two stages of the trial of $ \t p_{2, 2} = 0.0111 $ and $ \t p_{4,2}=0.0234 $. Then it is seen from Tables~\ref{table:finalbdriesJA} and \ref{table:finalbdriesJC} that all the p-value boundaries are crossed. Therefore $ H_{2} $ and $ H_{4} $ are both rejected under closed testing.

\section{Simulation Study} \label{sec:simulations}
\begin{figure}[htb]
	\centering
\input{graphsim}
 \caption{Graphical representation of  design for comparing PVcomb and CER in a trial with four treatment arms and two endpoints. The black arrow indicates the proportion of weight that is reallocated from the primary to the respective secondary hypotheses once the primary hypothesis is rejected. The blue dashed arrows indicate that the remaining weight is re-allocated to the other primary hypotheses. The orange arrows define that after the rejection of a secondary hypothesis, its weight is reallocated to the other primary hypotheses.   }\label{fig:simgraph1}
\end{figure}
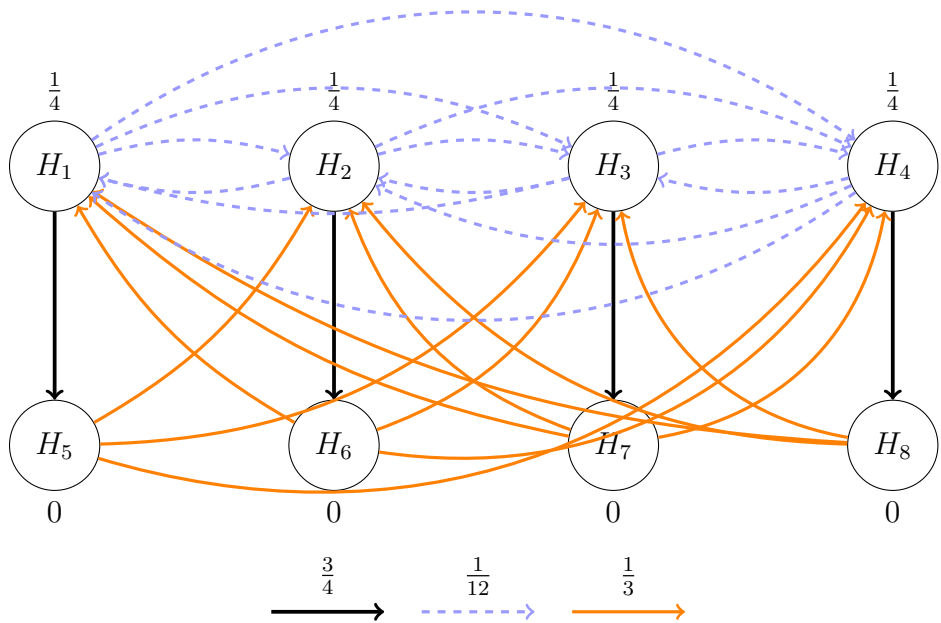

In order to compare the operating characteristics of the P-Value Combination (PVComb) and partial conditional error rate (CER) methods a simulation study was conducted.  We generalize the graph based test in Figure \ref{fig:schizo1} for a trial with four treatment arms that were compared in a pair-wise manner to a common control arm. Each treatment versus control comparison included two hypotheses, one primary and one secondary.  A closed test of the eight hypotheses was conducted over two stages with early rejection of efficacious hypotheses at the end of stage one based on the~\cite{gordon1983discrete} error spending function  (\ref{eq:ldof})  with $ t=0.5 $ and $ \gamma = 0.025 $. A graph based weighting strategy was adopted for testing all the intersection hypotheses of the closed test. The testing of intersection hypotheses exploited  known correlations among p-values with parametric tests, and reverted to Bonferroni based testing when the correlations were unknown, as described in Sections~\ref{sec:Combo} and \ref{sec:cer}.  The corresponding graph is depicted in Figure~\ref{fig:simgraph1}, where $ H_{1}, \ldots H_{4}  $ are the null hypotheses of no treatment effect for the four primary endpoints and $ H_{5}, \ldots H_{8} $ are the null hypotheses of no treatment effect for the corresponding secondary endpoints. Initially all the weight is distributed equally among the primary hypotheses. If a primary hypothesis is rejected, $3/4$ of its weight is transferred to the corresponding secondary endpoint. The remaining weight is equally distributed among the other primary hypotheses. If a secondary endpoint is rejected, its weight is transferred evenly to the remaining primary hypotheses.

We assume that the endpoints are normally distributed and we aim to test the null hypotheses that the difference in means in each endpoint between each treatment and control group are equal. The marginal p-values utilized in the hypothesis tests of each simulation were obtained from one-sided t-tests with degrees of freedom under the assumption of equal variance. The nominal familywise error rate was set to $2.5\%$.
Let $ \delta_{j}, j=1, \ldots 4 $  be the effect size of treatment $ j $ versus control for the primary endpoint. For this investigation we assumed that the effect size  of the secondary endpoint equals that of  the corresponding primary endpoint.  That is, $ \delta_{j}=\delta_{j+4}, j=1, \ldots 4 $. Additionally we assumed that there is a common standard deviation $ \sigma=1 $ for all treatment and control arms, for both endpoints. 

\begin{figure}[htb]
	\centering
	\includegraphics[width=1\linewidth]{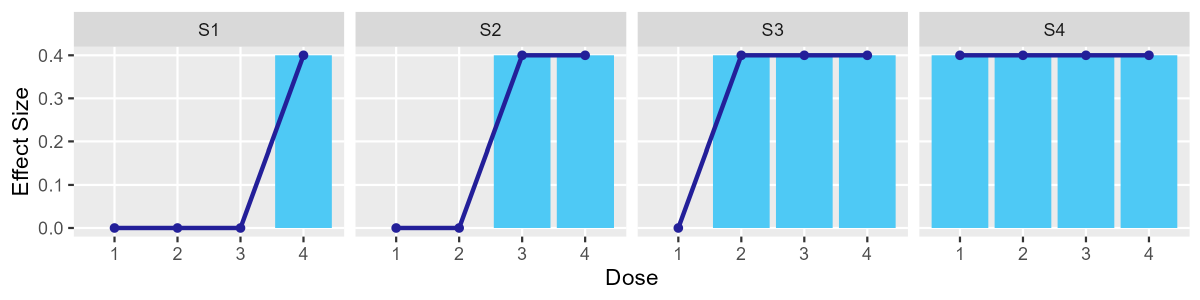}
	\caption{Graphical representation of  the effect size scenarios considered in the simulation study.}\label{fig:scenarios}
\end{figure}

We considered trials with a preplanned number of 100 subjects per arm and an interim analysis at 50\% of the enrollment for early rejection of statistically significant hypotheses and adaptive changes to the design. In the power simulations we assumed a correlation of 0.5 between the observations in the primary and secondary endpoint. To assess the type 1 error rate we additionally considered the case of no correlation and a correlation of 0.8. The simulations compared the disjunctive and conjunctive powers of the two methods over four alternative hypothesis scenarios and four decision rules for treatment selection at the end of stage one. The four alternative hypothesis scenarios, labelled S1 to S4 respectively, represent, one, two, three or four active treatment arms, with activity being defined as $ \delta_{j} =0.4 $, as shown in Figure~\ref{fig:scenarios}. For each scenario we conducted $500,000$ simulated trials for the Combo Method. For the CER Method, we conducted 100,000 stage one simulations and,  for each stage one result, we conducted 100 stage two simulations, and then aggregated the outcomes. As, for the CER method, the main computational effort is the computation of adjusted boundaries based on the first stage outcome, this approach is more efficient than a classical Monte Carlo simulation, simulating whole trials.

Four decision rules were utilized for dropping ineffective treatment arms at the end of stage one, based on marginal p-values. They ranged from extreme conservatism  to extreme aggression for dropping treatment arms and are thus titled as Conservative, Normal, Aggressive and Ultra Aggressive.  These rules are tabulated in Table~\ref{table:drules}. If a treatment arm was dropped at the end of stage one, the remaining subjects were reassigned to the arms going forward to stage two.

\begin{table}[htb]
\centering
\caption{Decision Rules for Dropping Treatment Arms}\label{table:drules}
\begin{tabular}{|l|l|}
\hline
\textbf{Decision Rule} & \textbf{Treatment Dropping Criterion} \\ \hline \hline
Conservative & $ \text{Drop if p-value} \geq 0.75 $ \\ \hline
Normal       & $ \text{Drop if p-value} \geq 0.5 $  \\ \hline
Aggressive   & $ \text{Drop if p-value} \geq 0.25 $ \\ \hline
Ultra Aggressive & Select only the treatment with the smallest p-value \\ \hline
\end{tabular}
\end{table}

\begin{figure}[htb]
	\centering
\includegraphics[width=1.0\linewidth]{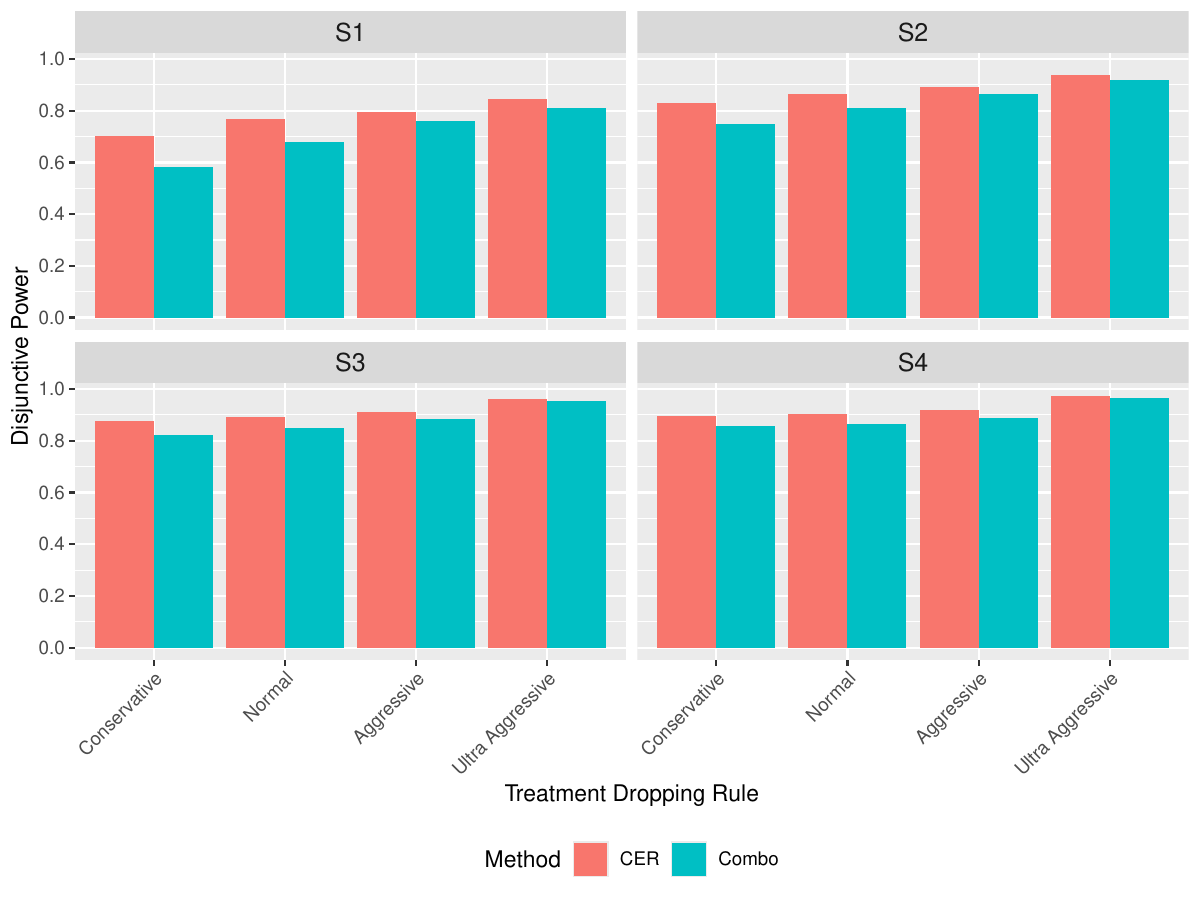}

\caption{ Disjunctive power of CER vs Combo by treatment dropping rule and scenario. }\label{fig:disjunct1}
\end{figure}

Figure~\ref{fig:disjunct1} compares the disjunctive power for the CER and Combo methods in four panels, one for each alternative hypothesis scenario. Within each panel, the comparisons are by treatment dropping decision rule.  Numeric simulation results are given in Table \ref{tab:power_table} in Appendix B.

\begin{figure}[htb]
	\centering
\includegraphics[width=1.0\linewidth]{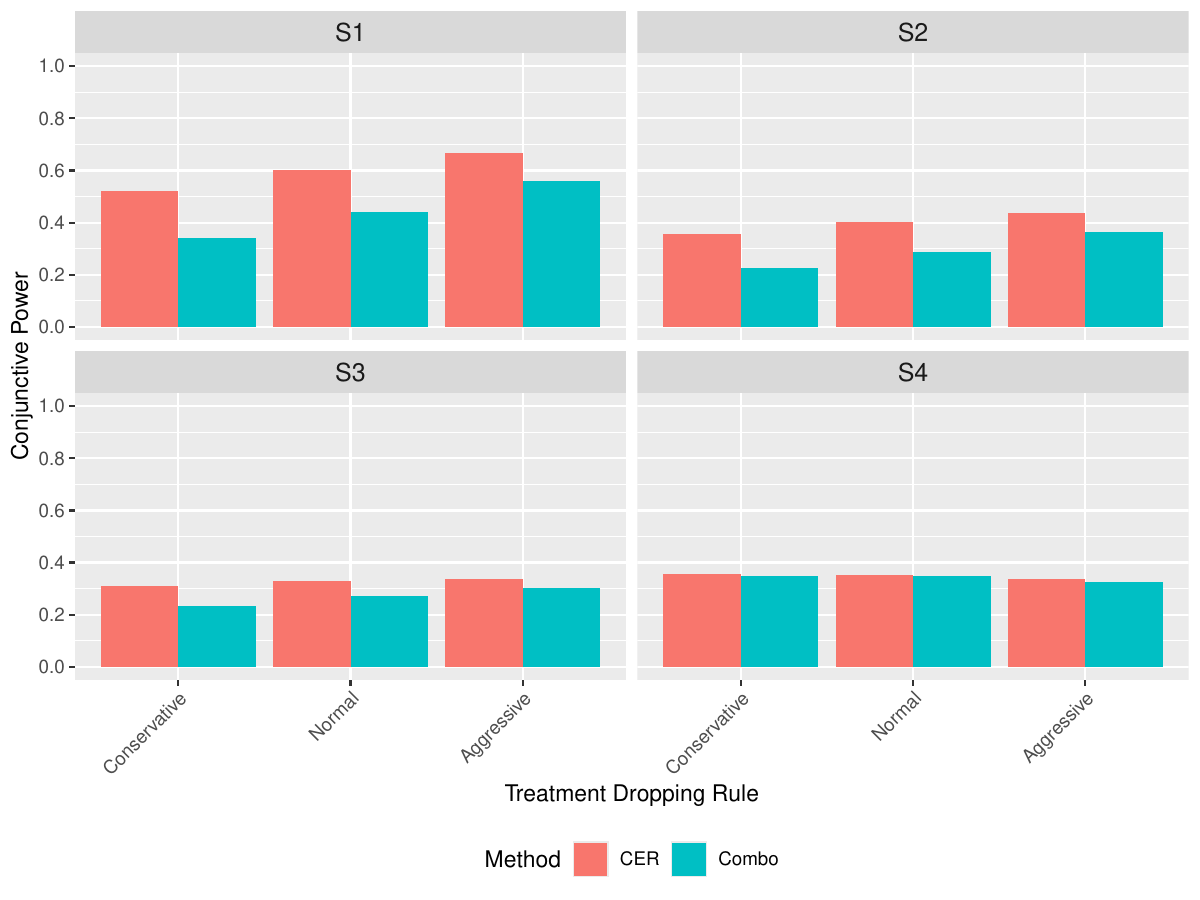}

\caption{ Conjunctive power of the CER vs Combo method by treatment dropping rule and scenario. For the Ultra Aggressive rule, that selects only a single arm for the second stage, no conjunctive power is reported. }\label{fig:conjunct1}
\end{figure}

The following consistent pattern emerges from an examination of these figures and table.
\begin{enumerate}
\item It is seen from Figure~\ref{fig:disjunct1} that the CER method has greater disjunctive power than the Combo in all considered scenarios. 
\item The gain in disjunctive power of CER over Combo within each alternative hypothesis scenario is greatest for the Conservative treatment dropping rule and diminishes with increasing aggressiveness for dropping losers.
\item The power gains of the CER method over Combo can be surprisingly large. For the Conservative treatment dropping rule the gain in disjunctive power ranges from 4\% to 12\%, while for the normal rule the range is 4\% to 9\%. 
\item Within each panel of Figure~\ref{fig:disjunct1} it is seen that the absolute disjunctive power for both methods increases with increasing aggressiveness of dropping treatments.
\item From a visual inspection of Figure~\ref{fig:simgraph1} it seems to be the case that the power gain of CER over Combo diminishes as the absolute power of both methods increases. This is to be expected since there is an upper bound on power. It does imply, however, that the settings in which the power gain is higher are precisely those in which it is most needed. 
\end{enumerate}

We also compared the two methods with respect to conjunctive power (see Figure \ref{fig:conjunct1} and Table \ref{tab:power_table} in  Appendix B). Again, in all considered scenarios the CER has a higher power with the largest differences with the conservative rule and when only few treatments are effective.

Finally, in order to verify FWER control, we conducted simulations under the global null hypotheses for all four decision rules under the assumption of no correlation, a correlation of 0.5, and a strong correlation of 0.8 among the primary and secondary endpoints (Table~\ref{table:FWERcontrol}).  The FWER  for both methods is controlled at the nominal level within the Monte Carlo error. However, in the considered scenarios the Combo method is substantially more conservative, with the exception of the Ultra Aggressive dropping rule. In addition, the FWER does not vary strongly between different strengths of correlation between the primary and secondary endpoint.

\begin{table}[htb]
\begin{center}
\caption{The estimated FWER of the two methods under the global null hypothesis. The standard error of the estimates is below 0.022  percentage points for the CER method and below 0.05 percentage points for the COMB method.}\label{table:FWERcontrol} 
\[
\begin{tabular}{|cl|cccc|}
\hline
&&\multicolumn{4}{c|}{FWER (in \%) by Decision Rules}\\ \hline
Correlation & Method & Conservative & Normal & Aggressive & Ultra \\
\hline
\multirow{2}{*}{0.0} & CER & 2.50 & 2.46 & 2.39 & 2.16 \\
& Combo & 1.12 & 1.21 & 1.55 & 2.38 \\
\hline
\multirow{2}{*}{0.5} & CER & 2.47 & 2.49& 2.42 & 2.24 \\
& Combo & 1.18 & 1.29 & 1.58 & 2.39 \\
\hline
\multirow{2}{*}{0.8} & CER & 2.48 & 2.46 & 2.40 & 2.31 \\
& Combo & 1.27 & 1.33 & 1.65 & 2.34 \\\hline
\end{tabular}
\]

\end{center}
\end{table}

\section{Discussion} \label{sec:conclusions}
This study has advanced the development of adaptive trial design in two ways. First, we have shown how to combine the graph-based methodology for testing multiple hypotheses with the group sequential methodology for early rejection of hypotheses. Second, we have demonstrated through a simulation study that the CER method for preserving the FWER of complex adaptive designs has greater power than the P-Value Combination method. This reinforces similar findings for the simpler multi-arm multi-stage single endpoint setting \citep{koenig2008adaptive,ghosh2020adaptive} and for enrichment designs \cite{sugitani2018flexible}. The large gains in power with CER compared to Combo under some scenarios and treatment dropping rules suggest that this method should be the preferred option for future trial design. We caution, however, that pre-trial simulations of all plausible scenarios and decision rules should always be conducted prior to settling on the choice of design. This includes simulations to verify strong control of the FWER.

We conjecture that the greater power of the CER over the Combo method is due to lack of consonance of the closed test (see \cite{hommel2007powerful}). For consonant tests,  rejection of the global intersection hypothesis implies rejection of at least one individual hypothesis. This does not necessarily hold for non-consonant tests. Therefore, non-consonant closed testing procedures are strictly conservative, as rejection of the intersection hypothesis at level $\alpha$ does not imply rejection of an  individual hypothesis in the closed test. This also results in a loss of power. The pre-planned test of the CER method, defined by the graph in Figure \ref{fig:simgraph1}, is consonant. Therefore, whenever the trial continues with all treatment arms to the second stage, for the CER method a rejection of the intersection hypothesis implies rejection of at least one individual hypothesis in the closed test. However, if treatment arms are dropped at the interim analysis, consonance is lost. This can be seen from Table \ref{table:FWERcontrol} that shows that for the CER method the FWER decreases for increasingly aggressive treatment dropping rules. For the Combo method in contrast, even if no treatments are dropped, the combination test is not consonant. Here, non-consonant test decisions are less frequent for more aggressive treatment dropping rules and therefore we also observe a better exhaustion of the significance level for these dropping rules. Overall, although for both,  CER and the Comb,  the adaptive test is in general not consonant the non-consonance can become much more severe for the Combo than the CER test. While the estimated FWER for the CER test is larger than 2.16\% in all considered scenarios, it can fall below 1.2\% for the Combo test (Table \ref{table:FWERcontrol}). We have included a a PDF file ({\tt appendix\_consonance.pdf}) in the Supplemental Material section on the Journal's site with  simple numerical examples of graph based tests that start out consonant at stage one and lose this property at stage two and also graph based tests that start out non-consonant at stage one and acquire consonance at stage two.

We also noted in Section~\ref{sec:simulations} that, for both methods, in the considered scenarios the absolute magnitude of the disjunctive power is lowest for the Conservative drop-the-loser rule and increases with increasing aggressiveness for dropping losers. This is readily explained by the fact that if a treatment arm is dropped at the interim analysis time point, the remaining sample size that was committed to that treatment arm at the design stage is re-allocated to the remaining arms that will be proceeding to stage two.

Although our worked example focused on comparing multiple treatments with multiple endpoints, the framework of graph-based multiple testing can be applied more generally. Thus, the multiple hypotheses could represent a population of interest and its underlying subgroups, possibly identified by biomarkers. In this setting, the methodology could be used for adaptive population designs. We will develop an application of this type in a subsequent paper.

We have confined our investigation to two-stage designs. The generalization to more than two stages is straightforward for the Combo method, since the same basic algorithm for combining incremental adjusted p-values with pre-specified weights can be applied. The generalization of the CER method, however, may require more care and is currently a topic for investigation. 

{\tt AdaptGMCP}, an R-package implementing the methods in this paper is available in the Supplemental Material section on the Journal site along with a PDF file ({\tt Illustrative Examples in Section 3.1.3 and 3.2.4.pdf}) that demonstrates its application to the worked examples in Sections 3.1.3 and 3.2.4 of the paper. {\tt AdaptGMCP} can  also be downloaded from \url{https://github.com/Cytel-Inc/AdaptiveGMCP}

\bibliographystyle{abbrvnat}
\bibliography{Refs2023}

\newpage

\begin{appendix}
\section{Appendix}
\renewcommand{\thetable}{A\arabic{table}} 
\setcounter{table}{0}

The  graph used in the simulation study in Section \ref{sec:simulations} is defined by the initial weight vector $(1/4,1/4,1/4,1/4,0,0,0,0)$ and the transition matrix
\begin{equation*}
G=\begin{bmatrix}
  0 & \frac{1}{12} & \frac{1}{12} & \frac{1}{12} & \frac{3}{4} & 0 & 0 & 0 \\
  \frac{1}{12} & 0 & \frac{1}{12} & \frac{1}{12} & 0 & \frac{3}{4} & 0 & 0 \\
  \frac{1}{12} & \frac{1}{12} & 0 & \frac{1}{12} & 0 & 0 & \frac{3}{4} & 0 \\
  \frac{1}{12} & \frac{1}{12} & \frac{1}{12} & 0 & 0 & 0 & 0 & \frac{3}{4} \\
  0 & \frac{1}{3} & \frac{1}{3} & \frac{1}{3} & 0 & 0 & 0 & 0 \\
  \frac{1}{3} & 0 & \frac{1}{3} & \frac{1}{3} & 0 & 0 & 0 & 0 \\
  \frac{1}{3} & \frac{1}{3} & 0 & \frac{1}{3} & 0 & 0 & 0 & 0 \\
  \frac{1}{3} & \frac{1}{3} & \frac{1}{3} & 0 & 0 & 0 & 0 & 0
\end{bmatrix}.
\label{Gconsgraph}
\end{equation*}

\section{Appendix}
\renewcommand{\thetable}{B\arabic{table}} 
\setcounter{table}{0} 

\begin{table}[ht]

\centering
\begin{tabular}{|c|c|lrrr|}
 \hline
Scenario & Dropping Rule & Method & Disjunctive & Conjunctive & FWER \\ 
\hline \multirow{8}{*}{S1} & \multirow{2}{*}{Conservative} & CER & 70.2 & 52.0 & 2.40 \\ 
   &  & Combo & 58.2 & 34.1 & 1.36 \\ 
   \cline{2-6}& \multirow{2}{*}{Moderate} & CER & 76.7 & 60.1 & 2.38 \\ 
   &  & Combo & 67.9 & 44.2 & 1.49 \\ 
   \cline{2-6}& \multirow{2}{*}{Aggressive} & CER & 79.6 & 66.8 & 2.35 \\ 
   &  & Combo & 76.1 & 56.0 & 1.73 \\ 
   \cline{2-6}& \multirow{2}{*}{Ultra Aggressive} & CER & 84.4 &  & 0.64 \\ 
   &  & Combo & 80.9 &  & 0.67 \\ 
  \hline \multirow{8}{*}{S2} & \multirow{2}{*}{Conservative} & CER & 83.1 & 35.6 & 2.19 \\ 
   &  & Combo & 75.0 & 22.7 & 1.63 \\ 
   \cline{2-6}& \multirow{2}{*}{Moderate} & CER & 86.6 & 40.2 & 2.22 \\ 
   &  & Combo & 80.9 & 28.9 & 1.77 \\ 
   \cline{2-6}& \multirow{2}{*}{Aggressive} & CER & 89.0 & 43.8 & 2.16 \\ 
   &  & Combo & 86.5 & 36.3 & 1.86 \\ 
   \cline{2-6}& \multirow{2}{*}{Ultra Aggressive} & CER & 93.6 &  & 0.28 \\ 
   &  & Combo & 92.1 &  & 0.28 \\ 
  \hline \multirow{8}{*}{S3} & \multirow{2}{*}{Conservative} & CER & 87.7 & 31.0 & 2.06 \\ 
   &  & Combo & 82.2 & 23.5 & 2.04 \\ 
   \cline{2-6}& \multirow{2}{*}{Moderate} & CER & 89.3 & 32.9 & 2.05 \\ 
   &  & Combo & 84.9 & 27.0 & 2.00 \\ 
   \cline{2-6}& \multirow{2}{*}{Aggressive} & CER & 91.2 & 33.5 & 1.89 \\ 
   &  & Combo & 88.6 & 30.1 & 1.86 \\ 
   \cline{2-6}& \multirow{2}{*}{Ultra Aggressive} & CER & 96.2 &  & 0.13 \\ 
   &  & Combo & 95.3 &  & 0.12 \\ 
  \hline \multirow{8}{*}{S4} & \multirow{2}{*}{Conservative} & CER & 89.7 & 35.5 &  \\ 
   &  & Combo & 85.6 & 34.9 &  \\ 
   \cline{2-6}& \multirow{2}{*}{Moderate} & CER & 90.4 & 35.3 &  \\ 
   &  & Combo & 86.4 & 34.7 &  \\ 
   \cline{2-6}& \multirow{2}{*}{Aggressive} & CER & 92.1 & 33.5 &  \\ 
   &  & Combo & 88.9 & 32.7 &  \\ 
   \cline{2-6}& \multirow{2}{*}{Ultra Aggressive} & CER & 97.4 &  &  \\ 
   &  & Combo & 96.7 &  &  \\ 
   \hline
\end{tabular}
\caption{Disjunctive and conjunctive power as well as the FWER in percent for the Conditional Error Rate (CER) and the combination test (Comb) method in the Scenarios S1, S2, S3, S4 and different rules to drop doses. For the Ultra Aggressive rule no conjunctive power is given. The standard error for the power estimates are below 0.1 percentage points, the standard errors for the FWER below 0.022 percentage points. In scenario S4 all null hypotheses are false and no type 1 error can occur.} 
\label{tab:power_table}
\end{table}

\end{appendix}
\end{document}

%% file: notation.tex
\usepackage{natbib}
\usepackage{array}
\usepackage{longtable}
\usepackage{setspace}
\usepackage{amsmath}
\usepackage{amsfonts}
\usepackage{amssymb}
\usepackage[utf8]{inputenc}
\usepackage[T1]{fontenc}
\usepackage{enumitem}
\usepackage{url}
\usepackage{amsbsy}
\usepackage{gastex}
\usepackage{xspace}
\usepackage{xifthen}
\usepackage{xcolor}
\usepackage{colortbl}
\usepackage{algorithm}
\usepackage{algorithmic}
\usepackage{multirow}

\usepackage{tikz}
\usepackage{tkz-graph}
\usetikzlibrary{positioning, shapes.geometric, calc}
\usepackage[colorinlistoftodos]{todonotes}
\usepackage{listings}
\usepackage{caption}

\renewcommand{\t}{\tilde}

\renewcommand{\P}{\mathcal{P}}
\renewcommand{\j}{\mathcal{J}}

\newcommand{\bct}{\begin{center}}
\newcommand{\ect}{\end{center}}
\newcommand{\bi}{\begin{itemize}}
\newcommand{\ei}{\end{itemize}}
\newcommand{\bed}{\begin{description}}
\newcommand{\eed}{\end{description}}
\newcommand{\be}{\begin{enumerate}}
\newcommand{\ee}{\end{enumerate}}
\newcommand{\bq}{\begin{quote}}
\newcommand{\eq}{\end{quote}}
\newcommand{\bqq}{\begin{quotation}}
\newcommand{\eqq}{\end{quotation}}
\newcommand{\beqn}{\begin{equation}}
\newcommand{\eeqn}{\end{equation}}

\raggedright
\raggedright

\newcommand{\adaptws}[1]{\boldsymbol{v}\ifthenelse{\isempty{#1}}{}{_{#1}}}

\newcommand{\adaptG}[1]{F\ifthenelse{\isempty{#1}}{}{_{#1}}}

\newcommand{\gf}[2]{
  g_{#1\ifthenelse{\isempty{#2}}{}{,#2}} 
}
\newcommand{\gfs}[1]{
  \left(\ifthenelse{\isempty{#1}}{\gf{ij}{}}{\gf{ij}{#1}}\right)_{i,j \in
    \ifthenelse{\isempty{#1}}{I}{#1}}
}

\newcommand{\gs}[2]{
  f_{#1\ifthenelse{\isempty{#2}}{}{,#2}} 
}
\newcommand{\gss}[1]{
  \left(\ifthenelse{\isempty{#1}}{\gs{ij}{}}{\gs{ij}{#1}}\right)_{i,j \in
    \ifthenelse{\isempty{#1}}{I}{#1}}
}

%% file: graphsim.tex
\begin{tikzpicture}[
    node distance=2.5cm,
    roundnode/.style={circle, draw=black, minimum size=1.2cm, inner sep=0pt}, 
    boldedge/.style={->, line width=.5mm, black}, % Thick black edges
    blueedge/.style={->, dashed, very thick, blue!40}, % Dashed blue edges
    rededge/.style={->, very thick, orange!99, bend right=20}, % Red thick edges bending the other way
    every loop/.style={min distance=20mm, looseness=10} % More space for loops
    ]

    \node[roundnode] (H1) at (0,0) {$H_1$};
    \node[above=0cm of H1] {$\frac{1}{4}$};

    \node[roundnode, right=of H1] (H2) {$H_2$};
    \node[above=0cm of H2] {$\frac{1}{4}$};

    \node[roundnode, right=of H2] (H3) {$H_3$};
    \node[above=0cm of H3] {$\frac{1}{4}$};

    \node[roundnode, right=of H3] (H4) {$H_4$};
    \node[above=0cm of H4] {$\frac{1}{4}$};

    \node[roundnode, below=of H1] (H5) {$H_5$};
    \node[below=0cm of H5] {$0$};

    \node[roundnode, below=of H2] (H6) {$H_6$};
    \node[below=0cm of H6] {$0$};

    \node[roundnode, below=of H3] (H7) {$H_7$};
    \node[below=0cm of H7] {$0$};

    \node[roundnode, below=of H4] (H8) {$H_8$};
    \node[below=0cm of H8] {$0$};

    \draw[boldedge] (H1) to (H5);
    \draw[boldedge] (H2) to (H6);
    \draw[boldedge] (H3) to (H7);
    \draw[boldedge] (H4) to (H8);

    \draw[rededge] (H5) to[bend right=15] (H2);
    \draw[rededge] (H5) to[bend right=25] (H3);
    \draw[rededge] (H5) to[bend right=35] (H4);
    
    \draw[rededge] (H6) to[bend left=15] (H1);
    \draw[rededge] (H6) to[bend right=25] (H3);
    \draw[rededge] (H6) to[bend right=35] (H4);
    
    \draw[rededge] (H7) to[bend left=15] (H1);
    \draw[rededge] (H7) to[bend left=25] (H2);
    \draw[rededge] (H7) to[bend right=35] (H4);
    
    \draw[rededge] (H8) to[bend left=15] (H1);
    \draw[rededge] (H8) to[bend left=25] (H2);
    \draw[rededge] (H8) to[bend left=35] (H3);

    \draw[blueedge] (H1) edge[bend left=15] (H2);
    \draw[blueedge] (H1) edge[bend left=25] (H3);
    \draw[blueedge] (H1) edge[bend left=35] (H4);

    \draw[blueedge] (H2) edge[bend left=15] (H1);
    \draw[blueedge] (H2) edge[bend left=15] (H3);
    \draw[blueedge] (H2) edge[bend left=25] (H4);

    \draw[blueedge] (H3) edge[bend left=15] (H1);
    \draw[blueedge] (H3) edge[bend left=15] (H2);
    \draw[blueedge] (H3) edge[bend left=15] (H4);

    \draw[blueedge] (H4) edge[bend left=35] (H1);
    \draw[blueedge] (H4) edge[bend left=25] (H2);
    \draw[blueedge] (H4) edge[bend left=15] (H3);

    \node[anchor=north] at ($(H7.south) - (1.8, .6)$) {%
        \begin{tikzpicture}[scale=1]
            \draw[boldedge] (0,0) -- ++(1.5,0) node[above=2pt, midway] { $\frac{3}{4}$};
            \draw[blueedge] (2,0) -- ++(1.5,0) node[above=2pt, midway,black] { $\frac{1}{12}$};
            \draw[rededge] (4,0) -- ++(1.5,0) node[above=2pt, midway,black] { $\frac{1}{3}$};
        \end{tikzpicture}
    };

\end{tikzpicture}

%% file: main.bbl
\begin{thebibliography}{25}
\providecommand{\natexlab}[1]{#1}
\providecommand{\url}[1]{\texttt{#1}}
\expandafter\ifx\csname urlstyle\endcsname\relax
  \providecommand{\doi}[1]{doi: #1}\else
  \providecommand{\doi}{doi: \begingroup \urlstyle{rm}\Url}\fi

\bibitem[Anderson et~al.(2022)Anderson, Guo, Zhao, and
  Sun]{anderson2022unified}
K.~M. Anderson, Z.~Guo, J.~Zhao, and L.~Z. Sun.
\newblock A unified framework for weighted parametric group sequential design.
\newblock \emph{Biometrical Journal}, 64\penalty0 (7):\penalty0 1219--1239,
  2022.

\bibitem[Bauer and Köhne(1994)]{bauer1994evaluation}
P.~Bauer and K.~Köhne.
\newblock Evaluation of experiments with adaptive interim analyses.
\newblock \emph{Biometrics}, pages 1029--1041, 1994.

\bibitem[Bretz et~al.(2009{\natexlab{a}})Bretz, Koenig, Brannath, Glimm, and
  Posch]{bretz2009adaptive}
F.~Bretz, F.~Koenig, W.~Brannath, E.~Glimm, and M.~Posch.
\newblock Adaptive designs for confirmatory clinical trials.
\newblock \emph{Statistics in medicine}, 28\penalty0 (8):\penalty0 1181--1217,
  2009{\natexlab{a}}.

\bibitem[Bretz et~al.(2009{\natexlab{b}})Bretz, Maurer, Brannath, and
  Posch]{bretz2009graphical}
F.~Bretz, W.~Maurer, W.~Brannath, and M.~Posch.
\newblock A graphical approach to sequentially rejective multiple test
  procedures.
\newblock \emph{Statistics in medicine}, 28\penalty0 (4):\penalty0 586--604,
  2009{\natexlab{b}}.

\bibitem[Bretz et~al.(2011)Bretz, Posch, Glimm, Klinglmueller, Maurer, and
  Rohmeyer]{bretz2011graphical}
F.~Bretz, M.~Posch, E.~Glimm, F.~Klinglmueller, W.~Maurer, and K.~Rohmeyer.
\newblock Graphical approaches for multiple comparison procedures using
  weighted bonferroni, simes, or parametric tests.
\newblock \emph{Biometrical Journal}, 53\penalty0 (6):\penalty0 894--913, 2011.

\bibitem[Burman et~al.(2009)Burman, Sonesson, and
  Guilbaud]{burman2009recycling}
C.-F. Burman, C.~Sonesson, and O.~Guilbaud.
\newblock A recycling framework for the construction of bonferroni-based
  multiple tests.
\newblock \emph{Statistics in medicine}, 28\penalty0 (5):\penalty0 739--761,
  2009.

\bibitem[Ghosh et~al.(2020)Ghosh, Liu, and Mehta]{ghosh2020adaptive}
P.~Ghosh, L.~Liu, and C.~Mehta.
\newblock Adaptive multiarm multistage clinical trials.
\newblock \emph{Statistics in Medicine}, 39\penalty0 (8):\penalty0 1084--1102,
  2020.

\bibitem[Hommel(2001)]{hommel2001adaptive}
G.~Hommel.
\newblock Adaptive modifications of hypotheses after an interim analysis.
\newblock \emph{Biometrical Journal: Journal of Mathematical Methods in
  Biosciences}, 43\penalty0 (5):\penalty0 581--589, 2001.

\bibitem[Hommel et~al.(2007)Hommel, Bretz, and Maurer]{hommel2007powerful}
G.~Hommel, F.~Bretz, and W.~Maurer.
\newblock Powerful short-cuts for multiple testing procedures with special
  reference to gatekeeping strategies.
\newblock \emph{Statistics in Medicine}, 26\penalty0 (22):\penalty0 4063--4073,
  2007.

\bibitem[Klinglmueller et~al.(2014)Klinglmueller, Posch, and
  Koenig]{klinglmueller2014adaptive}
F.~Klinglmueller, M.~Posch, and F.~Koenig.
\newblock Adaptive graph-based multiple testing procedures.
\newblock \emph{Pharmaceutical Statistics}, 13\penalty0 (6):\penalty0 345--356,
  2014.

\bibitem[Koenig et~al.(2008)Koenig, Brannath, Bretz, and
  Posch]{koenig2008adaptive}
F.~Koenig, W.~Brannath, F.~Bretz, and M.~Posch.
\newblock Adaptive dunnett tests for treatment selection.
\newblock \emph{Statistics in medicine}, 27\penalty0 (10):\penalty0 1612--1625,
  2008.

\bibitem[Lan and DeMets(1983)]{gordon1983discrete}
G.~K. Lan and D.~L. DeMets.
\newblock Discrete sequential boundaries for clinical trials.
\newblock \emph{Biometrika}, 70\penalty0 (3):\penalty0 659--663, 1983.

\bibitem[Magirr et~al.(2014)Magirr, Stallard, and Jaki]{magirr2014flexible}
D.~Magirr, N.~Stallard, and T.~Jaki.
\newblock Flexible sequential designs for multi-arm clinical trials.
\newblock \emph{Statistics in Medicine}, 33\penalty0 (19):\penalty0 3269--3279,
  2014.

\bibitem[Marcus et~al.(1976)Marcus, Eric, and Gabriel]{marcus1976closed}
R.~Marcus, P.~Eric, and K.~R. Gabriel.
\newblock On closed testing procedures with special reference to ordered
  analysis of variance.
\newblock \emph{Biometrika}, 63\penalty0 (3):\penalty0 655--660, 1976.

\bibitem[Maurer and Bretz(2013{\natexlab{a}})]{maurer2013memory}
W.~Maurer and F.~Bretz.
\newblock Memory and other properties of multiple test procedures generated by
  entangled graphs.
\newblock \emph{Statistics in medicine}, 32\penalty0 (10):\penalty0 1739--1753,
  2013{\natexlab{a}}.

\bibitem[Maurer and Bretz(2013{\natexlab{b}})]{maurer2013multiple}
W.~Maurer and F.~Bretz.
\newblock Multiple testing in group sequential trials using graphical
  approaches.
\newblock \emph{Statistics in Biopharmaceutical Research}, 5\penalty0
  (4):\penalty0 311--320, 2013{\natexlab{b}}.

\bibitem[M{\"u}ller and Sch{\"a}fer(2001)]{muller2001adaptive}
H.-H. M{\"u}ller and H.~Sch{\"a}fer.
\newblock Adaptive group sequential designs for clinical trials: combining the
  advantages of adaptive and of classical group sequential approaches.
\newblock \emph{Biometrics}, 57\penalty0 (3):\penalty0 886--891, 2001.

\bibitem[M{\"u}ller and Sch{\"a}fer(2004)]{muller2004general}
H.-H. M{\"u}ller and H.~Sch{\"a}fer.
\newblock A general statistical principle for changing a design any time during
  the course of a trial.
\newblock \emph{Statistics in medicine}, 23\penalty0 (16):\penalty0 2497--2508,
  2004.

\bibitem[Placzek and Friede(2019)]{placzek2019conditional}
M.~Placzek and T.~Friede.
\newblock A conditional error function approach for adaptive enrichment designs
  with continuous endpoints.
\newblock \emph{Statistics in Medicine}, 38\penalty0 (17):\penalty0 3105--3122,
  2019.

\bibitem[Posch and Bauer(1999)]{posch1999adaptive}
M.~Posch and P.~Bauer.
\newblock Adaptive two stage designs and the conditional error function.
\newblock \emph{Biometrical Journal: Journal of Mathematical Methods in
  Biosciences}, 41\penalty0 (6):\penalty0 689--696, 1999.

\bibitem[Posch et~al.(2011)Posch, Maurer, and Bretz]{posch2011type}
M.~Posch, W.~Maurer, and F.~Bretz.
\newblock Type i error rate control in adaptive designs for confirmatory
  clinical trials with treatment selection at interim.
\newblock \emph{Pharmaceutical statistics}, 10\penalty0 (2):\penalty0 96--104,
  2011.

\bibitem[Proschan and Hunsberger(1995)]{proschan1995designed}
M.~A. Proschan and S.~A. Hunsberger.
\newblock Designed extension of studies based on conditional power.
\newblock \emph{Biometrics}, pages 1315--1324, 1995.

\bibitem[Sugitani et~al.(2016)Sugitani, Bretz, and Maurer]{sugitani2016simple}
T.~Sugitani, F.~Bretz, and W.~Maurer.
\newblock A simple and flexible graphical approach for adaptive
  group-sequential clinical trials.
\newblock \emph{Journal of biopharmaceutical statistics}, 26\penalty0
  (2):\penalty0 202--216, 2016.

\bibitem[Sugitani et~al.(2018)Sugitani, Posch, Bretz, and
  Koenig]{sugitani2018flexible}
T.~Sugitani, M.~Posch, F.~Bretz, and F.~Koenig.
\newblock Flexible alpha allocation strategies for confirmatory adaptive
  enrichment clinical trials with a prespecified subgroup.
\newblock \emph{Statistics in Medicine}, 37\penalty0 (24):\penalty0 3387--3402,
  2018.

\bibitem[Xi et~al.(2017)Xi, Glimm, Maurer, and Bretz]{xi2017unified}
D.~Xi, E.~Glimm, W.~Maurer, and F.~Bretz.
\newblock A unified framework for weighted parametric multiple test procedures.
\newblock \emph{Biometrical Journal}, 59\penalty0 (5):\penalty0 918--931, 2017.

\end{thebibliography}
